\documentclass[varenna]{cimento}
\usepackage{amsmath, graphicx, amssymb}

\title{Experiments in Dilute Atomic Bose-Einstein Condensation}

\author{E.A. Cornell, J.R. Ensher, \atque C.E. Wieman}
\institute{JILA, University of Colorado and National Institute of
Standards and Technology and Department of Physics, University of
Colorado - Boulder, CO 80309-0440}

\begin{document}

\maketitle

\section{Introduction}
\subsection{Why BEC?}
In the month we began writing this paper, fourteen papers on the
explicit topic of Bose-Einstein condensation (BEC) in a dilute gas
appeared in the pages of {\it Physical Review}. Both theoretical
and experimental activity in BEC has expanded dramatically in the
three years since the first observation of BEC in a dilute atomic
gas.  One is tempted to ask, why? Why is there so much interest in
the field?  Is this burst of activity something that could have
been foreseen, in, say, 1990?

We feel the answer to this last question is yes: while
``interesting" is a quality which is impossible to define, there
was good reason to anticipate, even in 1990, that dilute-gas BEC
was going to be something worth investigating.  To see why, one
need only perform the following simple test.  Ask any physicist of
your acquaintance to compile a list of what he or she considers
the six most intriguing physical phenomena that occur at length
scales greater than a picometer and smaller than a kilometer.  In
our experience, such lists almost always include at least two of
the following three effects: superfluidity, lasing, and
superconductivity. These three topics all share two common
features: counter-intuitive behavior and macroscopic occupation of
a single quantum state. This then would have been our clue --- one
might have anticipated that dilute-gas BEC was to be interesting
because it shares the same underlying mechanism with the widely
appreciated topics of lasing, superfluidity, and
superconductivity, and yet is quite different from any of them:
Dilute-gas BEC is more amenable to microscopic analysis than
superfluidity, more strongly self-interacting than laser beams,
and it possesses a range of experimental observables which are
complementary to those of superconductivity.

This paper will not pretend to provide a thorough survey of the
status of dilute-gas BEC experiment.  Any attempt to do so at this
time would surely be obsolete by the time it appears in print ---
such is the pace of developments these days.  Rather, we will try
to cover the same pedagogical ground as the set of lectures
delivered by one of us (EAC) at the 1998 Enrico Fermi summer
school on Bose-Einstein condensation. The lectures were
coordinated with other experimenter-lectures at Varenna, and thus
were not meant to be comprehensive. This paper as well will be
selective in its focus.

An outline of this paper is as follows: We begin with an
introductory section including a review of ``ideal-gas" BEC theory
\cite{introref}. The second section presents a history of the
subject, and the third a forward-looking summary of BEC
technology. The fourth section introduces the important topic of
interactions in the condensate and surveys experimental work on
interactions. The fifth section discusses, mainly by reference to
published articles, experiments on condensate excitations.  The
sixth section is a brief essay on the meaning of the ``phase" of a
condensate. The seventh section surveys what is known about
heating processes in trapped atom samples, and examines the
implications for trapping very large samples. The eighth section
contains some general observations on atomic collisions relevant
to evaporation.

\subsection{BEC in an Ideal Gas}

It is conventional to begin a lecture on ideal Bose statistics
with a ritual chant, ``Imagine a system of $N$ indistinguishable
particles..." On occasion, however, it is worth pausing to reflect
on what an exotic thing it is to imagine, a system of
indistinguishable particles. One is willing enough to admit that
yes, two atoms can be so similar one to the other as to allow no
possibility of telling them apart. Why not? That is not so much to
concede.  The problem arises only when one is forced to confront
the physical implications of the concept of indistinguishable
bosons.  For example, if there are ten particles to be arranged in
two microstates of a system, the statistical weight of the
configuration \{ten particles in one state and zero in the other\}
is exactly the same as the weight as the configuration \{five
particles in one state, five in the other.\} We find this ratio of
$1:1$ disquieting and counter-intuitive.  The corresponding ratio
for distinguishable particles, ``Boltzmannons," would be $1:252$.

In the second \cite{Einstein25} of Einstein's two papers
\cite{Einstein25, Einstein24, Pais} on Bose-Einstein statistics,
Einstein acknowledges that in his statistics, ``The \ldots
molecules are not treated as statistically independent \ldots,"
and comments that the differences between distinguishable and
indistinguishable state counting ``express indirectly a certain
hypothesis on a mutual influence of the molecules which for the
time being is of a quite mysterious nature." The mysterious
statistics of indistinguishable bosons in turn mandates a variety
of exotic behavior, {\it e.g.} the famous enhanced probability for
scattering into occupied states and of course Bose-Einstein
condensation.

Why should the presence or absence of distinguishing tags on the
atoms so profoundly change their statistical behavior?  For the
particular case of photons, the problem may be more palatably
expressed in terms of fields, rather than particles.  When many
photons occupy a particular microstate, we talk about the electric
field amplitude in a particular mode.  It is relatively easy, at
least for the authors of this paper, to believe that having equal
field amplitudes in two adjacent modes is no more entropic than
having a large field in one mode and a small field in the other.
In some ways, this hasn't helped much. For as soon as we try to
apply the notion of fields back to atoms, we are left with a new
set of troubling notions -- can the probability amplitude for a
large collection of atoms ever really be like an electric field --
coherent, continuous, macroscopically occupied, macroscopically
observable?  The answer is ``absolutely yes!" Troubling or no,
this is exactly what a Bose condensate is about.

Proceeding conventionally, then, let us imagine a system of $N$
indistinguishable particles, distributed among the microstates of
a confining potential, such that any occupation number is
allowable.  The mean distribution, the Bose-Einstein distribution
(BED), may be derived in several different ways. (See for instance
\cite{Huang}.)  In the end one may always understand the BED as
the most random way to distribute a certain amount of energy among
a certain number of particles in a certain potential. The BED
gives the mean number of particles in the $i$th state as

\begin{equation}
n_{i}=\frac{1}{e^{(\epsilon_i-\mu)/kT}-1} \label{BED}
\end{equation}

\noindent where $\epsilon_i$ is the energy of a particle in the
$i$th state, $k$ is Boltzmann's constant, and $T$ and $\mu$ are
identified as the temperature and chemical potential,
respectively.  In the microcanonical understanding of the BED,
$\mu$ and $T$ are determined from the constraints on total number
$N$ and total energy $E$:

\begin{subequations} \label{sums}
    \begin{equation} \label{Nsum}
    N=\sum_{i}\frac{1}{e^{(\epsilon_{i}-\mu)/kT}-1}
    \end{equation}
    \begin{equation} \label{Esum}
        E=\sum_{i}\frac{\epsilon_i}{e^{(\epsilon_i-\mu)/kT}-1}
    \end{equation}
\end{subequations}

\noindent For large systems, the constraints may be written as
integrals

\begin{subequations} \label{ints}
    \begin{equation} \label{Nint}
        N=\int d\epsilon
        \frac{g(\epsilon)}{e^{(\epsilon-\mu)/kT}-1}
    \end{equation}
    \begin{equation} \label{Eint}
        E=\int d\epsilon \frac{\epsilon g(\epsilon)}{e^{(\epsilon-\mu)/kT}-1}
    \end{equation}
\end{subequations}

\noindent where $g(\epsilon)$ is the density of states in the
confining potential.

Equations \eqref{BED}-\eqref{ints} contain nearly all the ideal
gas physics of BEC.  The number of spatial dimensions, and the
effects of a confining potential, are all taken care of in the
power law of the density of states, $g(\epsilon)$.  The effects of
finite number, and of very asymmetric potentials
\cite{vanDruten97}, can be determined by using the sums rather
than the integrals to constrain $\mu$ and $T$. The critical
temperature, the ground-state occupation fraction and the specific
heat can all be calculated without difficulty.  Only number
fluctuations, which require a more careful consideration of the
underlying ensemble statistics, are left out of this picture. The
overall picture is sufficiently easy to understand that, if the
system truly \emph{were} an ideal gas, there would be little left
to study at this point.

As it actually has turned out, interactions between particles add
immeasurably to the richness of the system, and work in dilute-gas
BEC is in its early days.

\subsection{Some Real Systems} \label{realsystems}
Einstein's original conception of BEC was in an ideal gas, but the
first experiments in BEC were in superfluid helium, a liquid,
(which is to say a strongly correlated fluid, the opposite limit
from an ideal gas). The beautiful and startling experiments on
viscosity, vortices, and heat-flow in liquid helium, and the
ground-breaking theory those experiments inspired, more or less
defined the field of Bose-Einstein condensation for four decades
and more \cite{Hehistorynote}. The BEC concept has been put to use
in broader contexts over the years, however, in such diverse
topics as Kaons in neutron stars \cite{Brown:Kaons}, cosmogenesis,
and exotic superconductivity \cite{Ranninger:supercon}. Using the
term in its broadest meaning (``macroscopic number of bosons in
the same state") one needn't have a fluid of any sort -- lasers
and masers produce macroscopically occupied states of optical and
microwave photons, respectively.  For that matter, a portable
telephone, or even a penny-whistle produce macroscopic occupation
numbers of identical bosons (radio-frequency photons in one case,
acoustic-frequency phonons, in the other.)

In superfluid He-4, the bosons exist independently of the
condensate process.  The fermionic neutrons, protons, and
electrons that make up a He-4 atom bind to form a composite boson
at energies much higher than the superfluid transition
temperature.  There exists a broad family of physical systems,
however, in which the binding of the fermions to form composite
bosons, and the condensation of those bosons into a
macroscopically occupied state, occurs simultaneously. This is of
course  the famous BCS mechanism. Best-known for providing the
microscopic physical mechanism of superconductivity, the
Bose-condensed ``Cooper pairs" of BCS theory occur as well in
superfluid He-3 and may also be relevant to the dynamics of large
nuclei \cite{Iachello:nuclear} and of neutron stars.

Prior to the observation of BEC in a dilute atomic gas, the
laboratory system which most closely realized the original
conception of Einstein was excitons in cuprous oxide \cite{Lin93}.
Excitons are formed by pulsed laser excitation in cuprous oxides.
There exist meta-stable levels for the excitons which delay
recombination long enough to allow the study of a thermally
equilibrated Bose gas. The effective mass of the exciton is
sufficiently low that the BEC transition at cryogenic temperatures
occurs at densities which are dilute in the sense of the mean
particle spacing being large compared to the exciton radius.
Recombination events, which can be detected either electrically
\cite{Mysyrowicz79} or by collecting their fluorescence
\cite{Lin93}, are the main experimental observable.  The most
convincing evidence for BEC in this system is an excess of
fluorescence from ``zero-energy" excitons. In addition, anomalous
transport behaviour evocative of superfluidity has been observed
\cite{Fortin93}.

As an experimental and theoretical system, BEC in a dilute gas is
nicely complementary to the variety of BEC-like phenomena
described above. In terms of strength of interparticle
interaction, atomic gas BEC is intermediate between liquid helium,
for which interactions are so strong that they can not be treated
by perturbation theory, and photon lasers, for which
``interactions" (or nonlinearity) is small except in effectively
two-dimensional configurations. In terms of the underlying
statistical mechanics, atomic-gas BEC is most like He-4. Unlike in
superconductivity, the bosons are formed before the transition
occurs, and unlike in lasers, the particle number is conserved.
The properties of the phase transition, then, should most closely
resemble liquid helium.  Finally, in terms of experimental
observables, atomic-gas BEC is in a class entirely by itself. The
available laboratory tools for characterizing and manipulating
atomic-gas BEC are essentially orthogonal to those available for
excitonic systems or for liquid helium.

Although dilute-gas atomic BEC provides a nice foil to more
traditional BEC systems, it comes at a price: one is forced to
work deep in the thermodynamically forbidden regime.

\subsection{Why BEC in Dilute BEC is hard.}

\begin{figure}
\begin{center}
\includegraphics*[width = 0.75\linewidth]{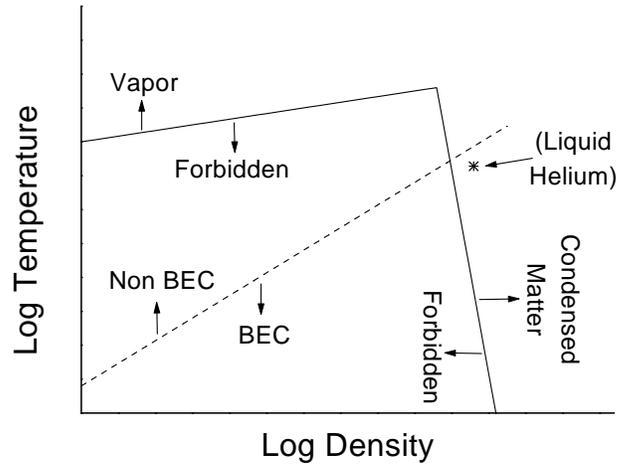}
\end{center}
\caption{Generic phase-diagram common to all atoms.  The dotted line shows
the boundary between non-BEC and BEC.  The solid line shows the boundary
between allowed and forbidden regions of the temperature-density space.
Note that at low and intermediate densities, BEC exists only in
the thermodynamically forbidden regime.} \label{phase}
\end{figure}

Figure \ref{phase} is a qualitative phase-diagram which shows the
general features common to any atomic system. At low density and
relatively high temperature, there is a vapor phase. At high
density there are various condensed matter phases. But the
intermediate densities are thermodynamically forbidden, except at
very high temperatures.  The Bose-condensed region of the $n-T$
plane is entirely forbidden, except at such high densities that
the equilibrium configuration of nearly all known atoms or
molecules is crystalline. (The existence of a crystal lattice
rules out a Bose condensate.)  The sole exception is helium, which
remains a liquid below the BEC transition.  Reaching BEC under
dilute conditions, say at densities ten or $100$ times lower than
conventional liquid helium, is as forbidden to helium as it is to
any other atom.

All the same, it is possible to do forbidden things, as long as
one doesn't attempt to do them for very long!  While making a
dilute-gas BEC in equilibrium is impossible, one can exploit
metastability to stray temporarily into the forbidden region. In
the absence of artificial nucleation sites, the process by which a
supersaturated atomic gas finds its way from a gas to a solid
begins with molecular recombination. Radiative recombination, by
which two atoms collide and emit a photon to stabilize a dimer
molecule, has a vanishingly small rate. At any reasonable density
the dominant channel is three-body recombination.  This then is
what makes BEC possible.  A gas of atoms can come into kinetic
equilibrium via two-body collisions, whereas it requires
three-body collisions to achieve chemical equilibrium ({\it i.e.}
to form molecules and thence solids.)  At sufficiently low
densities, the two-body rate will dominate the three-body rate,
and a gas will reach kinetic equilibrium, perhaps in a metastable
Bose-Einstein condensate, long before the gas finds its way to the
ultimately stable solid-state condition.

The need to maintain metastability usually dictates a more
stringent upper limit on density than does the desire to create a
dilute system.  Densities around $10^{20}$ cm$^{-3}$, for
instance, would be a hundred times more dilute than a
condensed-matter helium superfluid.  But creating such a gas is
quite impractical -- even at an additional factor of a thousand
lower density, say $10^{17}$ cm$^{-3}$, metastability times would
be on the order of a few ms; more realistic are densities on the
order of $10^{14}$ cm$^{-3}$. The low densities mandated by the
need to maintain long-lived metastability in turn challenge one's
ability to achieve the necessary low temperatures for BEC. The
critical temperature for ideal-gas BEC in a gas with mass $m$ is
related to density $n$ by

\begin{align}
 T_{\text{c,ideal}} = \frac{h^2}{2\pi m k}\biggl(\frac{n}{\zeta (3/2)}\biggr)^{2/3}\label{tc}
\end{align}

\noindent where $\zeta (z)$ is the Riemann zeta function and
$\zeta (3/2)\approx 2.612$. Realistically, one needs to be able to
achieve temperatures in the microkelvin scale or still lower,
which is a challenge indeed.

The desire to create a dilute BEC forces one to work in a
forbidden, metastable region of the $n-T$ plane. The need to
maintain metastability pushes one to still lower densities, and
low densities in turn supress the phase transition temperature to
temperatures that were all but unimaginable two decades ago. Thus,
making a dilute-gas BEC was an experimentally difficult thing to
do, and the history of the effort is worth reviewing.

\section{History}
   \subsection{Conceptual beginnings}
The notion of Bose statistics dates back to a 1924 paper in which
Satyendranath Bose used a statistical argument to derive the
black-body radiation spectrum \cite{Bose24}. Albert Einstein
extended the statistical model to include systems with conserved
particle number \cite{Einstein25, Einstein24}. The result was
Bose-Einstein statistics. Particles that obey Bose-Einstein
statistics are called bosons, and today it is known that all
particles with integer spin (and only those particles) are bosons.
Einstein immediately noticed a peculiar feature of the
distribution: at low temperature, it saturates. ``I maintain that,
in this case, a number of molecules steadily growing with
increasing density goes over in the first quantum state (which has
zero kinetic energy) while the remaining molecule separate
themselves according to the parameter value $A=1$ [in modern
notation, $\mu = 0$] \ldots A separation is effected; one part
condenses, the rest remains a `saturated ideal gas.' "
\cite{Einstein25,Pais} Thus began the concept of Bose-Einstein
condensation.

BEC has not always been a particularly reputable character. In the
decade following Einstein's papers, doubts were cast on the
reality of the model \cite{Gavroglu95}. Fritz London and L. Tisza
\cite{London38, Tisza38} resurrected the idea in the mid 1930s as
a possible mechanism underlying superfluidity in liquid helium 4.
London's view was either disbelieved or else felt to be not
particularly illuminating. Certainly the influential helium theory
papers of the 50s and 60s make little or no mention of BEC.
\cite{Landau41, Feynman55, Pitaevski61}. The authors of the
present review are not well-versed in the history of that era, but
it is evident that sometime in the intervening years the bulk of
expert opinion has shifted. Experiment and theory (neutron
scattering \cite{Sokol:helium} and path-integral Monte Carlo
simulations \cite{Ceperley95}, respectively) now support the idea
that the microscopic physics underlying superfluidity is a
zero-momentum BEC. Due to interactions, only about 10\% of the
helium atoms participate in the condensate, even at temperatures
so low that empirically 100\% of the liquid appears to be in the
super-fluid state.

\subsection{Spin-polarized hydrogen}
Efforts to make a dilute BEC in an atomic gas were spurred by
provocative papers by Hecht \cite{Hecht59} and Stwalley and
Nosanov \cite{Stwalley76}. They argued on the basis of the quantum
theory of corresponding states that spin-polarized hydrogen would
remain a gas down to zero temperature, and thus would be a great
candidate for making an weakly interacting BEC. A number of
experimental groups \cite{Silvera80, Hardy82, Hess83, Johnson84}
in the late 70s and early 80s began work in the field.
Spin-polarized hydrogen was first stabilized by Silvera and
Walraven in 1980 \cite{Silvera80}, and by the the mid-80s
spin-polarized hydrogen had been brought within a factor of $50$
of condensing \cite{Hess83}. These experiments were performed in a
dilution refrigerator, in a cell whose walls were coated with
superfluid liquid helium.  A radiofrequency (rf) discharge
dissociated hydrogen molecules, and a strong magnetic field
preserved the polarization of the resulting atoms. Individual
hydrogen atoms can thermalize with a superfluid helium surface
without becoming depolarized. The atoms were compressed using a
(conceptually) simple piston-in-cylinder arrangement
\cite{Bell86}, or inside a helium bubble \cite{Sprik85}.
Eventually the helium surface became problematic, however. If the
cell is made relatively cold, the surface density of hydrogen
atoms becomes so large that they undergo recombination there. If
the cell is too hot, the volume density of hydrogen necessary for
BEC becomes so high that, before that density can be reached, the
rate of three-body recombination becomes too high \cite{Hess86}.

\subsection{Laser cooling and the ascendancy of alkalis}
Contemperaneous with (but quite independent from) the hydrogen
work, an entirely different kind of cold-atom physics was
evolving.  The remarkable story of laser cooling has been reviewed
elsewhere \cite{Arimondo92, Chu98, CCT98, Phillips98} but we
mention some of the highlights in compressed form below. The idea
that laser light could be used to cool atoms was suggested in
early papers from Wineland and Dehmelt \cite{Wineland75}, from
H{\"a}nsch and Schawlow \cite{Hansch75}, and from Letokhov's group
\cite{Letokhov68}. Early optical force experiments were performed
by Ashkin \cite{Bjorkholm78}. Trapped ions were laser-cooled at
the University of Washington \cite{Neuhauser78} and at the
National Bureau of Standards (now NIST) in Boulder
\cite{Wineland78}. Atomic beams were deflected and slowed in the
early 80s \cite{Andreev81, Prodan85, Ertmer85}. Optical molasses
was first studied at Bell Labs \cite{Ashkin85}. Measured
temperatures in the early molasses experiments were consistent
with the so-called Doppler limit, which amounts to about $300$
microkelvin in most alkalis. Coherent stimulated forces of light
were studied \cite{Aspect86,Martin87}. The dipole force of light
was used to confine atoms as well \cite{Chu86}. In 1987 and 1988
there were two major advances that became central features of the
method of creating BEC. First, a practical spontaneous-force trap,
the Magneto-Optical trap (MOT) was demonstrated \cite{Raab87}, and
second, it was observed that under certain conditions, the
temperatures in optical molasses are in fact much colder than the
Doppler limit \cite{Lett88, Dalibard88, Chu88}. These were heady
times in the laser-cooling business. With experiment yielding
temperatures mysteriously far below what theory would predict, it
was clear that we all lived under the authority of a munificent
God. The MOT had generated high densities in trapped gases.
Perhaps still colder temperatures and still higher densities were
due to arrive shortly. Indeed, it seemed reasonable enough, at the
time, to speculate that the methods of laser cooling and trapping
might soon lead directly to BEC!

\subsection{Cold reality: limits on laser cooling}

It didn't happen that way, of course. By 1990 it was clear that
there were fairly strict limits to both the temperature and
density obtainable with laser cooling.  Theory caught up with
experiment and showed that the sub-Doppler temperatures were due
to a combination of light-shifts and optical pumping that became
known as Sysiphus cooling.  Random momentum fluctuations from the
rescattered photons limit the ultimate temperature to about a
factor of ten above the recoil limit \cite{Dalibard89}. The
rescattered photons are also responsible for a density limit --
the light pressure from the reradiated photons gives rise to an
effective inter-atom repulsive force \cite{Walker90}.  Momentum
fluctuations and trap loss arising from light-assisted collisions
limited temperature and density as well \cite{Vigue86}. The
product of the coldness limit and the density limit works out to a
phase-space density of about $10^{-5}$, which is five orders of
magnitude too low for BEC.

Since 1990, advances in laser cooling and trapping have allowed
both the temperature \cite{Kasevich94, CCT95} and the density
\cite{Ketterle93} limit to be circumvented. But it seems that in
most cases higher densities have been won at the cost of high
temperatures, and lower temperatures have been achievable only in
relatively low density experiments. The peak phase-space density
in laser cooling experiments has increased hardly at all in the
decade since 1989 \cite{lighttraps}.

The alkali-atom work of the 1980s included another development,
however, which was to have a large impact on BEC work. Sodium
atoms were trapped in purely magnetic traps \cite{Pritchard83,
Migdall85, Bagnato87}.

\subsection {Evaporative cooling and the return of hydrogen}
Harold Hess from the MIT Hydrogen group realized the significance
that magnetic trapping had for their BEC effort. Atoms in a
magnetic trap have no contact with a physical surface and thus the
surface-recombination problems could be circumvented. Moreover,
thermally isolated atoms in a magnetic trap were the perfect
candidate for evaporative cooling. In a remarkable paper Hess laid
out in 1986 most of the important concepts of evaporative cooling
of trapped atoms \cite{Hess86}. Let the highest energy atoms
escape from the trap, and the mean energy, and thus the
temperature, of the remaining atoms will decrease.  In a dilute
gas in an inhomogeneous potential, decreasing temperature in turn
means decreasing occupied volume. One can actually increase the
density of the remaining atoms even though the total number of
confined atoms decreases. The Cornell University Hydrogen group
also considered evaporative cooling \cite{Lovelace85}. By 1988 the
MIT group had implemented these ideas and had demonstrated that
the method was as powerful as anticipated. In their best
evaporative run, they obtained, at a temperature of $100$ $\mu$K,
a density only a factor of five too low for BEC \cite{Doyle91b}.
Further progress was limited by dipolar relaxation, but perhaps
more fundamentally by loss of signal-to-noise, and the need for a
more accurate means of characterizing temperature and density in
the coldest clouds \cite{Doyle91}. Evaporative work was also
performed by the Amsterdam group \cite{Luitin93}.

Both the MIT and the Amsterdam group began constructing laser
sources capable of accessing optical transitions in atomic
hydrogen, reasoning that with the powerful tool of laser
spectroscopy they would be better positioned to understand and
surpass the low temperature limits \cite{Luitin93, Sandberg93,
Cesar96}. The Amsterdam group performed some laser cooling on
atomic hydrogen \cite{Hijmans89, Setija}.

The hydrogen evaporation results made a big impression on the JILA
alkali group. It seemed to us that a hybrid approach combining
laser cooling and evaporation had an excellent chance of working.
Evaporation from a magnetic trap seemed like a very appealing way
to circumvent the limits of laser cooling. Laser cooling could
serve as a pre-cooling technology, replacing the dilution
refrigerator of the hydrogen work.  With convenient lasers in the
near-IR, and with the good optical access of a room-temperature
glass cell, detection sensitivity could approach single-atom
capability.  The elastic cross-section, and thus the rate of
evaporation, should almost certainly be larger in an alkali atom
than it is in hydrogen.  Encouraged by these thoughts (and by
other collisional considerations, see sections \ref{collisions}
and \ref{collisionscaling} (below) the JILA group set out (See
Ref. \cite{Monroe90} and a presentation at the 1991 Varenna summer
school \cite{Monroe92}) to combine the best ideas of alkali and of
hydrogen experiments, in an attempt to see BEC in an alkali gas.

\subsection{Hybridizing MOT and evaporative cooling techniques}
In a sense efforts to hybridize optical cooling with magnetic
trapping are as old as atomic magnetic trapping itself. The
original NIST sodium magnetic trap \cite{Migdall85} and the first
Ioffe-Pritchard (IP) trap \cite{Bagnato87} were loaded from
Zeeman-tuned optical beam-slowers.  Modern alkali BEC apparatuses,
however, can trace their conceptual roots through a series of
devices built at JILA during an era beginning in late 1989 and
continuing into the early 90s \cite{Monroe90, Monroe92, Cornell91,
Monroe93, Anderson94, Myatt96, Newbury95}. As things stood at the
end of the 1980s, optical cooling on the one hand and magnetic
trapping on the other were both somewhat heroic experiments, to be
undertaken only by advanced and well-equipped AMO laboratories.
The prospect of trying to get both working, and working well, on
the same bench and on the same day was daunting.

\begin{figure}
\begin{center}
\includegraphics*[width = 0.75\linewidth]{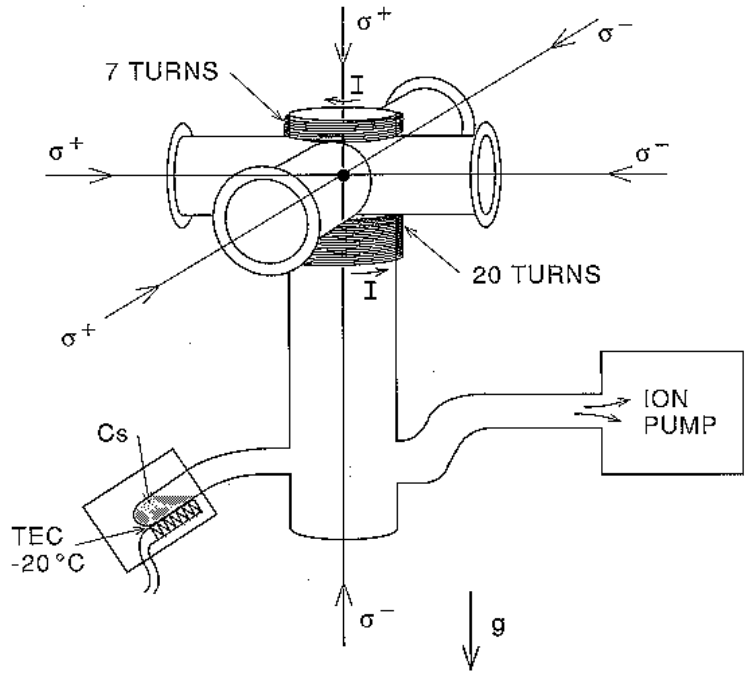}
\end{center}
\caption{The glass vapor cell and magnetic coils used in early
JILA efforts to hybridize laser cooling and magnetic trapping
\protect{\cite{Monroe90}}.  The glass tubing is 2.5 cm in
diameter. The Ioffe current bars have been omitted for clarity.}
\label{vaporcell}
\end{figure}

\begin{figure}
\begin{center}
\includegraphics*[width = 0.75\linewidth]{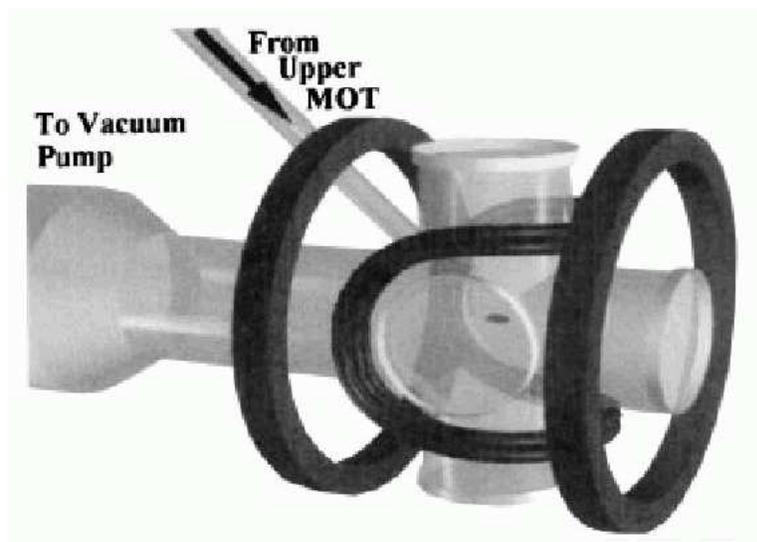}
\end{center}
\caption{Glass vapor cell and magnetic field coils used in mixed-BEC
studies \protect{\cite{Myatt97}} and ongoing Feshbach resonance
experiments \protect{\cite{Roberts98}}.} \label{cellmyatt}
\end{figure}

The JILA vapor-cell MOT, with its superimposed IP trap (figure
\ref{vaporcell}), represented a much-needed technological
simplification and introduced a number of ideas that are now in
common use in the hybrid trapping business: \cite{Monroe90,
Monroe92}: i) Vapor-cell (rather than beam) loading; ii)
fused-glass rather than welded-steel architecture; iii) extensive
use of diode lasers; iv) magnetic coils located outside the
chamber; v) over-all chamber volume measured in cubic centimeters
rather than liters; vi) temperatures measured by imaging an
expanded cloud; vii) magnetic-field curvatures calibrated \emph{in
situ} by observing the frequency of dipole and quadrupole
(sloshing and pulsing) cloud motion; viii) the basic approach of a
MOT and a magnetic trap which are spatially superimposed (indeed,
which often share some magnetic coils) but temporally sequential;
and ix) optional use of additional molasses and optical pumping
sequences inserted in time between the MOT and magnetic trapping
stages. In the early experiments \cite{Monroe90, Cornell91,
Monroe92, Monroe93} a number of experimental issues came up that
continue to confront all BEC experiments: the importance of
aligning the centers of the MOT and the magnetic trap; the
density-reducing effects of mode-mismatch; the need to account
carefully for the (previously ignored) force of gravity; heating
(and not merely loss) from background gas collisions; the
usefulness of being able to turn off the magnetic fields rapidly;
the need to synchronize many changes in laser status and magnetic
fields together with image acquisition. At the time the design was
quite novel, but by now it is almost standard. It is instructive
to note how much a modern, IP-based BEC device (figure
\ref{cellmyatt}) resembles its ancestor (figure \ref{vaporcell}).

\subsection{Collisional concerns} \label{collisions}
From the very beginning, dilute-gas BEC experiments had to
confront the topic of cold collisions.  Even before evaporation
was considered, when cooling was simply a matter of conduction to
the chamber walls, the rate of three-body decay determined the
lifetime of samples of spin-polarized hydrogen samples. With the
advent of evaporation, there was a demand for understanding
several different collisional processes -- elastic collisions,
dipolar relaxation, three-body recombination, and (to a lesser
extent) spin-exchange.

Atomic collisions at very cold temperatures is now a major branch
of the discipline of AMO physics, but in the 1980s there was
almost no experimental data, and what there was came in fact from
the spin-polarized hydrogen experiments \cite{Gillaspy89}. There
was theoretical work on hydrogen from Shlyapnikov and Kagan
\cite{Kagan81, Kagan84}, and from Silvera and Verhaar
\cite{Langendijk86}. An early paper by Pritchard
\cite{Pritchard86} includes estimates on low-temperature
collisional properties for alkalis. His estimates were
extrapolations from room-temperature results, but in retrospect
several were surprisingly accurate.

Experiments on cold collisions between alkali atoms were initially
all performed on light-induced collisions in molasses and MOTs
\cite{Prentiss88, Sesko89, Gould88}. The earliest ultra-cold
ground-state on ground-state measurements were based on
pressure-shifts in the cesium clock transition \cite{Gibble93} and
thermalization rates in magnetically trapped Cs \cite{Monroe93},
Rb \cite{Newbury95} and Na \cite{Davis95a}. S-wave collisions were
observed directly in cesium \cite{Gibble95}. Eventually experiment
and theory of excited state collisions, in particular
photoassociative events, yielded so much information on atom-atom
potentials that ground-state cross-sections could be extracted
\cite{Thorsheim87, Miller93, Lett93, Gardner95, McAlexander95,
Abraham95}.  By the time BEC was observed in rubidium, sodium, and
lithium, there existed, at least at the 20\% level, reasonable
estimates for the respective elastic scattering length of each
species.  Evaporation efforts on all three species were initially
begun, however, under conditions of relatively large uncertainty
concerning elastic rates, and near-ignorance on inelastic rates.

In 1992 the JILA group came to realize that dipolar relaxation in
alkalis should in principle not be a limiting factor.  As
explained in section \ref{evapcriteria} below, collisional scaling
with temperature and magnetic field is such that, except in
pathological situations, the problem of good and bad collisions in
the evaporative cooling of alkalis is reduced to the ratio of the
elastic collision rate to the rate of loss due to imperfect
vacuum; dipolar relaxation and three-body recombination can be
finessed.  It was reassuring to move ahead on efforts to evaporate
with the knowledge that, while we were essentially proceeding in
the dark, there were not as many monsters in the dark as we had
originally imagined. Of course, even given successful evaporative
cooling, there was still the question of the sign of scattering
length, which must be positive to ensure the stability of a large
condensate.  The JILA group had sufficient laser equipment,
however, to trap either $^{85}$Rb, $^{87}$Rb, or $^{133}$Cs. Given
the ``modulo" arithmetic that goes into determining a scattering
length, it seemed fair to treat the scattering lengths of the
three species as statistically independent events, and the chances
then of Nature conspiring to make all three negative were too
small to worry about \cite{potentialthoughts}.

\subsection{Increasing elastic collisions per background loss}
A simple MOT, with operating parameters adjusted to maximize
collection rate, will confine atoms at a density limited by
reradiated photon pressure to a relatively low value
\cite{Walker90}, with correspondingly low collision rate after
transfer to the magnetic trap.

\subsubsection{Multiple Loading} Early efforts to surmount the density limit
included a multiple-loading scheme pursued at JILA
\cite{Cornell91}. Multiple MOT-loads of atoms were launched in
moving molasses, optically pumped into an untrapped Zeeman level,
focused into a magnetic trap, then optically repumped into a
trapped level. The repumping represented the necessary
dissipation, so that multiple loads of atoms could be inserted in
a continuously operating magnetic trap.  In practice, each step of
the process involved some losses and the final result hardly
justified all the additional complexity.  It is interesting to
note, however, that multiple transfers from MOT to MOT
\cite{Gibble95, Myatt96}, rather than from MOT to magnetic trap,
is a technique currently in widespread practice.

\subsubsection{Adiabatic compression} Also in wide, current use is the
practice of magnetic compression. After loading into a magnetic
trap, the atoms can be compressed by further increasing the
curvature of the confining magnetic fields. This is a technique
which is available to experiments which use MOTs to load the
magnetic trap, but not to experiments which use cryogenic loading
(in the latter, the magnetic fields are usually already at their
maximum values during loading, in order to maximize capture.) This
method is discussed in \cite{Monroe92} and was implemented first
in early ground-state collisional work \cite{Monroe93}.

\subsubsection{Enhancing MOT density}
An important advance came from the MIT Sodium group in 1992, when they
developed the Dark-spot MOT \cite{Ketterle93}.  A shadow is
arranged in the repumping light, such that atoms that have already
been captured and concentrated in the middle of the trap are
pumped to an internal state that is relatively unperturbed by
close neighbors. In effect, the MOT is divided into two spatial
regions, one for efficient collection, and one for efficient
compression and storage. A related technique is to modulate the
MOT parameters temporally, rather than spatially. This approach
has been dubbed CMOT (Compressed MOT) \cite{Petrich94}. A very
thorough study of related MOT behavior came out of a
British-French collaboration \cite{Townsend95}.

\subsubsection{Reducing background loss}
Since the most important figure of merit in the evaporation
business in the collision rate compared to the trap lifetime, one
can do almost as much good improving vacuum as one can improving
MOT density. Early machines for cooling atomic beams were
relatively dirty, by the standards of modern UHV practice. Except
for the Pritchard group's cryogenic trap \cite{Bagnato87},
confinement times in optical trapping experiments were usually on
the order of two or three seconds. Improved atomic beam design
\cite{Ketterle95, Tollett95} and the adoption of modern UHV
practice has made 100 s lifetimes standard.  An experiment at JILA
saw a MOT lifetime of 1000 s \cite{Anderson94}. Cryogenic MOTs
have had lifetimes of greater than 3600 s \cite{Libbrecht95}.

\subsection{Forced evaporation}
Cooling by evaporation is a process found throughout nature.
Whether the material being cooled is an atomic nuclei or the
Atlantic ocean, the rate of natural evaporation, and the minimum
temperature achievable, are limited by the particular fixed value
of the work-function of the evaporating substance.  In
magnetically confined atoms, no such limit exists, because the
``work function" is simply the height of the lowest point in the
rim of the confining potential.  Hess pointed out \cite{Hess86}
that by perturbing the confining magnetic fields, the
work-function of a trap can be made arbitrarily low; as long as
favorable collisional conditions persist, there is no lower limit
to the temperatures attainable in this forced evaporation.
Pritchard \cite{Pritchard89} pointed out that evaporation could be
performed more conveniently if the rim of the trap were defined by
an rf-resonance condition, rather than simply by the topography of
the magnetic field; experimentally, his group made first use of
position-dependent rf transitions to selectively transfer
magnetically trapped sodium atoms between Zeeman levels and thus
characterized their temperature \cite{Martin88}.  From 1993-1995,
a number of other experimental groups launched their own efforts
to create BEC in alkali species using the hybrid
laser-cooling/evaporation approach. Progress accelerated and by
summer of 1994 three groups had announced the successful use of
evaporation to increase the phase-space density of trapped alkalis
\cite{Petrich95, Adams95, Davis95a}.  By the DAMOP meeting in
Toronto, in May of 1995, the number of groups with clear evidence
of evaporative cooling had increased to four \cite{Hulet95}. The
Rice, JILA, and MIT alkali groups were all seeing significant
increases in phase-space density. The JILA alkali group returned
from Toronto with a shared impression that ``there is no time left
to fiddle around" \cite{Anderson95b}.

\subsection {Magnetic trap improvements}
Each of the  groups reporting significant evaporation at Toronto,
had recently implemented a major upgrade in magnetic trap
technology. In a harmonic trap, the collision rate after adiabatic
compression scales as the final confining frequency squared
\cite{Monroe92}. Clearly, there was much to be gained by building
a more tightly confining magnetic trap, but the requirement of
adequate optical access for the MOT, along with engineering
constraints on power dissipation, make the design problem
complicated. When constructing a trap for weak-field seeking
atoms, with the aim of confining the atoms to a spatial size much
smaller than the size of the magnets, one would like to use linear
gradients.  In that case however one is confronted with the
problem of the minimum in the magnitude of the magnetic fields
(and thus of the confining potential) occurring at a local zero in
the magnetic field. This zero represents a ``hole" in the trap, a
site at which atoms can undergo Majorana transitions
\cite{Majorana31} and thus escape from the trap. If one uses the
second-order gradients from the magnets to provide the
confinement, there is a marked loss of confinement strength. This
scaling is discussed in reference \cite{Petrich95}. The three
experimental groups each solved the problem differently. The Rice
group built a trap with permanent magnets \cite{Tollett95} which
can be stronger than electromagnets. The MIT and JILA groups each
built traps with linear gradients, and then ``plugged the hole,"
in the MIT case with a beam of blue-detuned (repulsive) light
\cite{Davis95a}, and in the JILA case with a time-dependent dither
field \cite{Petrich95}. As it turned out, all three methods worked
well enough.  In Spring of 1995 (less than three weeks after the
DAMOP meeting), the JILA group saw BEC in rubidium 87, and by
Autumn of 1995 all three groups had either observed BEC or else
had preliminary evidence for it \cite{Anderson95, Davis95b,
Bradley95}.

If one is willing to accept tight confinement in two directions
only, it turns out that the original IP design, operated with a
very low bias field, is more than adequate to produce BECs. In
1996, BEC was observed in second-generation traps at both MIT
\cite{Mewes96a} and JILA \cite{Myatt97}. The fields were in the IP
configuration; additional coils were used to nearly null the
central magnetic field.

\subsection{Imaging techniques }

The fact that a wealth of quantitative information can be
extracted from simple spatial images of the atom cloud is one of
the features that most distinguishes BEC in dilute atomic gas from
other BEC experiments, such as excitons and liquid helium. It is
worth mentioning then some issues in imaging trapped atoms. Early
imaging  of magnetically trapped atoms was done by collecting
fluorescence  \cite{Monroe90}.  While acceptable for large, dilute
clouds, fluorescence imaging does not work well when the clouds
get small and dense because of optical thickness and motional
blurring. Absorption imaging minimizes motional blurring because
the mean spontaneous force from the absorbed light is directly
along the line of sight (assuming good optical alignment.) The
presence of magnetic fields in the trap can make absorption images
more difficult to interpret, although the MIT Sodium group has
shown that with careful modeling even images taken of atoms in
full-strength quadrupole magnetic fields can be correctly
understood \cite{Davis95a}. The more fundamental problem is that
as the clouds get colder, they get both smaller and denser.  Two
issues arise. First, the cloud can become smaller than the
resolution limit of the optics, which makes it easy to
misinterpret images \cite{Bradley95, Cornell98b} . Second, very
dense clouds can give rise to ``lensing," in which the real part
of the index of refraction becomes as important as the absorption.
In the early experiments, the solution was to turn off the fields
suddenly and let the cloud expand before imaging. The expansion
decreases the optical thickness and increases the effective
resolution.  For the original JILA observation, it was necessary
to magnetically support the atoms against gravity in order to
allow the atoms to expand sufficiently without dropping out of the
imaging field-of-view. More recently {\it in situ} imaging
techniques have been demonstrated to be accurate (see Section
\ref{imaging} below).


\section{Survey of BEC technology, present and future}

The list of groups currently observing BEC, and the basic
technology each uses (Table \ref{BECtechnology}) will doubtless be
outdated by the time this article appears in print, but it is
interesting to note the  variations on a theme. There
are many obvious generalizations to make, but most of them have
exceptions. In this section we survey the range of existing
technology and speculate on what the future may bring.

\subsection{Magnetic (and other) Traps}

1. As a rule, these are of the IP configuation. But TOP traps
continue to be built, and the IP traps have yet to zero-in on a
standard configuration. As of this writing, the eleven groups
producing BEC with IP fields are using nine different coil
geometries.  Still more exotic trap designs have been proposed
\cite{Shapiroxx}.  It seems likely that over the next few years
there will be something of a shake-out in coil design, with a
small subset of today's designs being found to be adequate to
cover the needs of most experiments.

2. As a rule, the coils are water-cooled. But the MIT hydrogen
trappers use superconducting coils, and the Orsay group uses so
little power that they can air-cool.

3. It seems likely that evaporative cooling to BEC can eventually
be accomplished by purely optical means, either in FORTs
\cite{Adams95}, or their near-dc cousins, the QUESTs
\cite{Takekoshi95, Takekoshi96}. Already, the MIT group has
confined condensates in purely optical traps
\cite{Stamper-Kurn98a}, although the clouds are still precooled in
magnetic traps. Still more exotic traps have been proposed,
including boxes fashioned from blue-detuned evanescent optical
waves or from planes of tiny magnetic domains \cite{Hinds98}.

\begin{table}
\begin{tabular}{|c|c|c|l|} \hline

Place &Atom& Collector/Cooler & Magnetic Trap \\[0pt]\hline

JILA \cite{Anderson95}      & $^{87}$Rb  & MOT/MOT & i) TOP

\\ & & & ii)IP (baseball) \\\hline

MIT \cite{Davis95b}       & $^{23}$Na  & Beam/MOT & IP
(cloverleaf)    \\ & & &(hybrid optical) \\

\hline Rice \cite{Bradley95}      & $^7$Li  & Beam/Molasses & IP
(permanent magnet) \\

\hline Rowland Institute \cite{Hau98}  & $^{23}$Na & Beam/MOT & IP
(four-dee) \\

\hline Yale (was Stanford) \cite{Anderson98}      & $^{87}$Rb &
Vapor/MOT     & TOP \\

\hline UTexas \cite{Han98}    & $^{87}$Rb & Beam/MOT      & TOP \\

\hline Konstanz \cite{Ernst98}  & $^{87}$Rb & MOT/MOT       & IP
(bars) \\

\hline M{\"{u}}nchen \cite{Esslinger98}   & $^{87}$Rb & MOT/MOT
& IP (3 coils) \\

\hline NIST       & $^{23}$Na & Beam/MOT      & TOP (transverse)
\\ Gaithersburg \cite{Deng98} & & &\\

\hline Paris \cite{Dalibard98}    & $^{87}$Rb  & MOT/MOT       &
IP (3 coils) \\

\hline MIT \cite{Fried98}      & H  & Dilution Refrigerator/ & IP
(Superconducting) \\         &    & He-II Surface & \\

\hline Orsay \cite{Aspect98} &$^{87}$Rb&Beam/MOT & IP
(Pole-Piece)\\

\hline Hannover \cite{Ertmer98}&$^{87}$Rb&Beam/MOT&IP(Cloverleaf)
\\

\hline Otago \cite{Wilson98}  & $^{87}$Rb & MOT/MOT     & TOP \\

\hline Sussex \cite{Boshier98}& $^{87}$Rb & MOT/MOT     & IP
(Baseball)\\

\hline
\end{tabular}
\caption{Current BEC technology at a glance --- Where, What and
How} \label{BECtechnology}
\end{table}

\subsection{Atomic species}
A majority of groups are currently working in rubidium 87, with
most of the remainder using sodium 23. But the Rice group has had
success in lithium 7 and the hydrogen trappers recently have seen
condensation \cite{Fried98}. There is considerable experimental
effort around the world on cesium \cite{Arndt97, Soding98,
Guery-Odelin98, Arlt98, Arimondo98}, rubidium 85 \cite{Roberts98},
and potassium 39 and 41 \cite{Fort98, Prevedelli98} and on the
fermionic isotopes lithium 6 \cite{Houbiers98} and potassium 40
\cite{Cataliotti98, DeMarco98}. To the best of our knowledge, all
sodium condensates to date have been formed in the $F=1$ state.
There seems to be some suggestion that the collisional properties
of the $F=2, m_F=2$ state are unfavorable for condensation. With
the recent successes of buffer-gas loading of molecules and
non-alkali atoms \cite{Kim97, Weinstein98, Weinstein98b} it seems
likely that almost any atomic species with a ground-state magnetic
moment, and a great many molecular species, can be magnetically
trapped. The success of sympathetic cooling \cite{Myatt97} lifts
most of the requirements on collisional properties. A species with
recalcitrant collisional properties can be trapped together with,
and cooled sympathetically by, a ``good evaporating" species such
as $^{87}$Rb; sympathetic cooling requires a far smaller good/bad
collision ratio to succeed. Thus the number of different atomic
and molecular species that could eventually be cooled to the BEC
transition may be in the hundreds.

\subsection{Precooling and compression}

Evaporative cooling can work only if the sample is initially cold
and dense enough to permit tight confinement and large collision
rates.  By far the most popular approach is to collect, cool, and
compress atoms in a MOT \cite{Raab87}, then rapidly turn off the
MOT and turn on the fields of the magnetic trap immediately around
it \cite{Monroe90}. The density of the MOT is often enhanced by a
darkspot \cite{Ketterle93}, or by modulating the detuning and
field gradients of the MOT just before transfer to the magnetic
trap \cite{Petrich94}. Sometimes a stage of optical cooling is
included after the MOT magnetic fields have been turned off and
before the magnetic trap has been turned on \cite{Monroe90}, but
for traps with large number, this often provides only marginal
benefit.  The MOT is a useful but, as it has turned out, hardly an
essential tool. The Rice group, which cannot rapidly reconfigure
their magnetic fields from MOT to IP trap, uses a beam-loaded
molasses and dispenses with a MOT altogether.  The hydrogen
trapping groups have never needed molasses cooling of any kind to
prepare atoms for magnetic trapping and evaporative cooling --
they rely on thermalization with a dilution-refrigerated He-II
surface. In the future, buffer-gas loading techniques \cite{Kim97}
may rival MOTs as the primary precooling method. Optical cooling
may also continue even after the atoms are in their final
evaporation trap. Optical cooling has been shown to work in
magnetic traps \cite{Helmerson92, Newbury95} and in optical dipole
traps \cite{Boiron98, Lee98}.

\subsection {MOT loading}
Among the experiments that do use MOTs, there are variants in the
methods for loading them. The simplest approach is to collect
atoms in the MOT directly from back-ground thermal vapor
\cite{Monroe90, Anderson95}, but one encounters a dilemma --
Should one operate at high vapor pressure, so as to fill the MOT
quickly with large numbers of atoms, but then suffer rapid loss
during evaporation, or should one operate at low vapor pressure,
so as to have longer time available for evaporation, but then live
with slow, meager TOP trap fills?  The original JILA apparatus
found a workable range of compromise pressures, with the help of a
darkspot used to suppress collisional loss from the MOT during the
200-second fill-time \cite{Anderson94}.  The desire for larger
condensates and faster repetition rates has prompted most groups
to abandon the simplicity of a single vapor cell in favor of more
elaborate loading schemes.  Roughly half of the groups fill their
MOTs with Zeeman-slowed beams \cite{Prodan85}; most of the
remainder use a double-MOT scheme -- a ``dirty" MOT in a
relatively high pressure region cools atoms which are transferred
to a ``science MOT" at UHV \cite{Gibble95, Myatt96, Burt97}. In
the future, LVIS \cite{Lu96} or funnel-type beam generators
\cite{Ito97} may replace the Zeeman slowed beams. A range of other
possibilities exist: one could devise a way to rapidly modulate
the vapor pressure of the desired species, initially high for MOT
collection, then low for evaporation. Alternatively, it has been
shown also to be possible to spatially separate the MOT from the
magnetic trap, and accumulate many MOT loads in a single magnetic
trap \cite{Cornell91}; perhaps such a technique could eventually
find use.  In yet another direction, the group at JILA is
investigating  mechanical means for moving atoms from high-density
collection regions to low-density evaporation regions.

\subsection{Chamber Design}
As a rule, the chambers tend to be small glass cells with magnetic
coils wound outside of the vacuum, but the more traditional
conflat-based stainless-steel cans with vacuum-compatible coils
inside are in use as well, for instance at the Rowland Institute
and at the University of Texas. Most likely stainless-steel
chambers will continue to find use in certain applications -- in
cryogenic systems, for instance, or in experiments for which the
condensate is to be deposited on or scattered off of samples. Such
experiments may benefit from the greater flexibility of
reconfigurable steel chambers.

\subsection{Imaging technique} \label{imaging}
Imaging the cloud after ballistic expansion has the advantage of
being relatively easy to do, but the technique has its drawbacks.
Some traps are difficult to turn off rapidly enough to ensure that
no unintentional (and uncharacterized) work is done on the
expanding gas.  Also, although in most cases the expansion process
is relatively easy to model \cite{Holland96, Castin96}, it is an
additional and sometimes unwelcome link in the chain of inference
that connects an image with the underlying physical behavior.
Finally, post-expansion imaging is not consistent with acquiring
multiple images of the same cloud. The alternative to
post-expansion imaging, imaging \emph{in situ}, has the previously
mentioned drawbacks that the cloud can be small compared to the
resolution limit, and that the cloud can be so optically thick as
to give rise to lensing systematics. If sufficient care is taken,
quantitative information can be extracted from an absorption image
even when the object is smaller than the resolution limit
\cite{Bradley97a, Bradley97b}, or when the object is very
optically thick \cite{Hau98}. The problems of optical thickness
can be circumvented altogether with imaging methods based on
far-detuned probe beams sensitive solely to the real part of index
of refraction: Polarization rotation \cite{Bradley97a}, and
dark-field and phase-contrast imaging \cite{Andrews96}.

For work requiring very high spatial resolution, or high
signal-to-noise, expansion followed by absorption imaging will
continue to be a useful method.  For work in which the (relatively
thin) thermal component is of primary interest, \emph{in situ}
absorption imaging will probably be preferable. For studies of
time-dependent behavior (as long as they do not require very high
spatial resolution) {\it in situ} phase-contrast methods are
probably best.

\section{Effects of interactions}
As discussed in Section \ref{realsystems} above, the topic of
Bose-Einstein condensation develops its full richness only after
interactions are included. Most recent discussions of interactions
in a $T=0$ condensate begin with the Gross-Pitaevski (GP)
equation:

\begin{align}
i \hbar\frac{\partial\Psi}{\partial t} = \biggl(
-\frac{\hbar^{2}\nabla^2}{2 m} + V_{\text{ext}} +
\frac{4\pi\hbar^{2}a}{m}|\Psi|^2\biggr)\Psi\label{gp}
\end{align}

\noindent where $a$ is the 2-body scattering length,
$V_{\text{ext}}$ is the external potential of the trap and $\Psi$
is the order parameter. In a dilute gas at zero temperature, the
density at any point is space and time is just $|\Psi|^2$. For an
instructive comparison of several quite distinct routes to arrive
at this ``starting point" see \cite{Esry97c}. For recent surveys
of the status of the theory in this area, see \cite{DalfovoXX,
Parkins98, Griffin98, Fetter98, Burnett98}. We will review here
the experimental studies to date on interactions. All published
experiments on BEC to date have shown signs of these interactions,
and these experiments have tested the theoretical predictions to a
greater or lesser degree. For convenience we will discuss
excitations in a separate section.

\subsection{Energy and size}
In an ideal gas, the statistical mechanics of Bose condensation
ensures that the majority of the atoms will accumulate in the
lowest energy state of the confining potential. In particular, in
a harmonic trap, the atoms accumulate in the (0,0,0) state of the
3-D harmonic oscillator, with a zero-point energy given by
$(\hbar/2)(\omega_x + \omega_y + \omega_z)$ per atom. The onset of
inter-particle interactions does not change the basic behavior of
BEC: atoms still accumulate in the lowest energy state. The
difference is only that the spatial extent, and total energy, of
the interacting ground-state is larger than its ideal gas
predecessor.

The energy of the condensate is proportional to the square of the
size of its time-of-flight image.  Interactions perturbed even the
earliest images of $2000$-atom condensates \cite{Anderson95,
Holland96}, and are the dominant factor in determining the size of
larger condensates \cite{Mewes96a}.  Once the number of atoms in a
condensate becomes sufficiently large, the interaction energy and
the external potential energy (second and third terms, left side
of eq. \eqref{gp} above) dominate the quantum kinetic energy (the
first term).  In solving for the ground-state of eq. \eqref{gp} it
then becomes convenient to ignore the kinetic term, an approach
known as the Thomas-Fermi (TF) approximation.  Once in the TF
regime, the size of the condensate scales as $N^{1/5}$ and its
energy as $N^{2/5}$ \cite{Lovelace87}. The energy scaling behavior
in the TF limit was confirmed with expanded image data from the
MIT group \cite{Mewes96a}. In the intermediate regime between the
ideal-gas limit and the TF limit, all three terms in eq.
\eqref{gp} are significant, and one expects to see cross-over
behavior in the energy dependence. This was observed in
expanded-image data at JILA \cite{Jin96b, Holland97}.  The JILA
data also verified that the coefficient in the interaction term of
eq. \eqref{gp} was within $20\%$ of the predicted value.

\begin{figure}
\begin{center}
\includegraphics*[width = 0.75\linewidth]{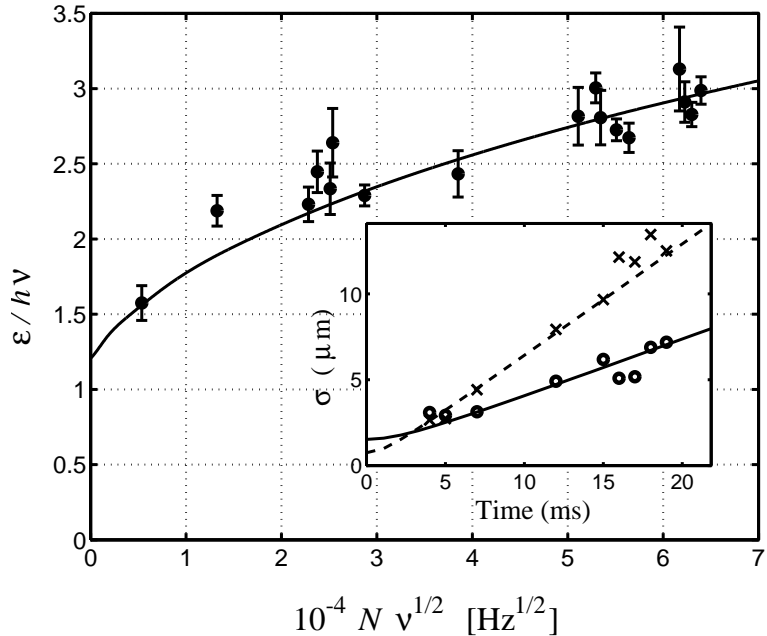}
\end{center}
\caption{Comparison of the measured condensate energy
(\protect{$\bullet$}) to the prediction from mean-field theory
(solid line) as a function of interaction strength. By
non-adiabatically releasing the condensate from the magnetic trap,
the kinetic and interaction energies of the trapped atoms become
the kinetic energy of the released and expanding atom cloud. The
kinetic energy of expansion is obtained from a sequence of cloud
widths measured at different expansion times. The inset shows
experimental widths in the horizontal (\protect{$\circ$}) and
vertical (\protect{$\times$}), compared to the mean-field
predictions (dashed and solid lines), for the data point at
\protect{$10^{-4}N\nu^{1/2}=0.53$ Hz$^{1/2}$}. Used with
permission of M. Holland from ref. \protect{\cite{Holland97}}.}
\label{interaction}
\end{figure}

Quantitative checks on the \emph{in situ} size of the interacting
condensate have been performed at MIT \cite{Andrews97b} and the
Rowland institute \cite{Hau98}, and are in good agreement with
theory.

In addition to total energy, and \emph{in situ} size of the
condensate, eq. \eqref{gp} also yields predictions on the aspect
ratio of the expanded cloud and on the detailed shape of the
condensate density profile, in the trap or expanded, (in the TF
limit, the density profile is an inverted paraboloid). See
\cite{DalfovoXX} for a review of this theory. In one
high-precision measurement, the aspect ratio of an expanded cloud
was measured to be in agreement with theory at the $2\%$ level
\cite{Matthews98}. In the presence of noise and various image
distortions, it can be difficult to quantify how well a ``shape"
agrees with theory, but no zero-$T$ measurements we are aware of
\cite{Castin96} show any major discrepancies with the basic
theory.

\subsection{Two-species condensates}
In thinking about experiments in condensate mixtures, it is
easiest to imagine mixtures of two different elements, rubidium
and sodium, for instance \cite{Law97}.  As it turns out,
experiments with two-species condensates have to date mainly been
performed in mixtures of two hyperfine states of $^{87}$Rb
\cite{Myatt97, Hall98b, Hall98c, Cornell98, Cornell98c}.

The two-component rubidium work has been recently reviewed
\cite{Cornell98}, so we will not discuss it extensively here. A
range of two-fluid behavior, including mutual repulsion, component
separation, relative-center-of-mass oscillation and damping, and
residual steady-state overlap have been observed. Qualitatively,
the mutual interaction behavior can be accounted for in the
framework of two, coupled, GP equations \cite{Ho96b, Esry97b,
Law97, SinatraXX}. Quantitatively, this model has not been tested
to any degree of accuracy.

It is worth noting that the rf output-coupler experiment of MIT
\cite{Mewes97} was also sensitive to interactions between
different spin-states of the same atomic species. The crescent
shape of the out-coupled pulse of atoms arose from the repulsion
between the out-coupled $F=1, m_F=0$ atoms, and the still-trapped
$F=1, m_F=-1$ atoms \cite{Zhang98}.

Experiments have also been recently performed in a
multiple-component system in which spontaneous interconversion of
components is an important process in the system's dynamics. These
are the $F=1$ spinor condensates of the MIT sodium group. This
work is reviewed in another article in this volume
\cite{Ketterle98}.

\subsection{Negative scattering length} Bose condensates composed
of atoms with negative scattering lengths pose a particular
challenge to theory. In a homogenous system, such a condensate can
not exist, because it is unstable to small density perturbations
--- for wavelengths larger than a critical value, excitation
frequencies become negative.  However, if a condensate is confined
in a potential such that its size is smaller than the critical
wavelength, it is stable (or at least metastable) to density
fluctuations.  Given a particular atomic species and a particular
confining potential, there is a critical number of condensate
atoms. If the condensate grows beyond that number, it will
implode.  Currently the only published experimental data on this
phenomenon are from the Rice Lithium group \cite{Bradley97a}. At
present they do not have the stability or sensitivity to watch a
single condensate approach the critical value and then collapse.
Rather, they collect many images and show that they never observe
condensates larger than a certain size, a size consistent with the
predicted ``implosion limit."

In future work, improved stability and statistical analysis of
observed cloud sizes may shed more light on the issue of
metastability \cite{Hulet98}. Alternatively, experiments utilizing
atomic Feshbach resonances may be able to study implosion dynamics
more directly.

\subsection{Finite temperature}
The theory of Bose condensation at nonzero temperature
\cite{Griffin98, Fetter98, Burnett98} is a more subtle topic than
$T=0$ BEC. After all, at $T=0$ the phenomenon of ``Bose
condensation," in the sense of a large fraction of the particles
occupying the lowest available orbital, would occur even in
Maxwell-Boltzmann statistics \cite{Holzmann98}. In the decades
following the original Einstein papers, it was even believed the
phase transition predicted in 1925 \cite{Einstein25} would not
survive a proper treatment of interactions at finite temperature
\cite{Gavroglu95, Holzmann98}. Experimentally, we now know that
indeed the phase transition does occur \cite{Ensher96} at close to
the (non-zero) temperature predicted in \cite{Einstein25}, and
that the associated coherent behavior \cite{Andrews97a, Burt97,
Ketterle97} survives the effects of interactions and finite
temperature. Beyond that, there has been relatively few critical
experimental tests of finite-$T$ theory. There are several
published experimental studies on the effect of temperature on
excitations, but these data, as discussed in section
\ref{finitetempexc} below, are taken in a regime for which
theoretical modeling is particularly difficult. There are however
solid theoretical predictions for the effects of finite-$T$
interactions on several static and thermodynamic properties:

\subsubsection{Critical temperature}  Interactions should affect
the temperature at which the condensate first appears in two
different ways. First, the mean-field repulsion arising from the
atoms concentrated at the lowest point in the confining potential
will reduce the density there; the net effect is to reduce the
critical temperature relative to what it would be for the same
number of non-interacting atoms in the same external potential.
See for instance \cite{Bagnato87b}.  Second, many-body effects
among the confined particles, effects which can be described as a
break-down of the dilute limit, are predicted to modestly
\emph{increase} the critical temperature \cite{Krauth96,
Bijlsma96}.  So far there has been no decisive experimental
observation of either effect. The most precise measurement of
$T_{\text{c}}$ to date \cite{Ensher96} quoted an error of about
$5\%$, which was comparable to the expected effect of
interactions. Improvements in the accuracy of measurements of $N$,
$N_o$, and $T$ will probably allow an unambiguous confirmation
that interactions \emph{affect} $T_{\text{c}}$, but a quantitative
sorting out of the relative effects of mean-field and many body
interactions will be an enormous experimental challenge
\cite{Houbiers97, Holzmann98b}.

\begin{figure}
\begin{center}
\includegraphics*[width = 0.75\linewidth]{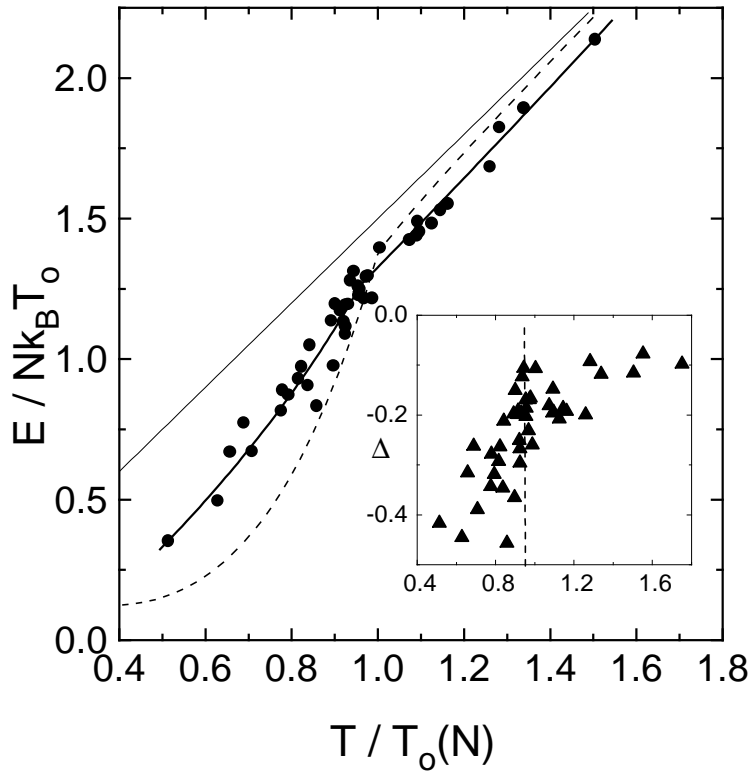}
\end{center}
\caption{The measured scaled energy per particle of a trapped Bose
gas is plotted vs. the temperature scaled by the critical
temperature, $T_{\text{o}}$, for a non-interacting (ideal) gas of
bosons in a 3-D harmonic oscillator. The straight, solid line is
the energy for a classical, ideal gas, and the dashed line is the
predicted energy for a finite number of ideal bosons
\protect\cite{Grossman95, Ketterle96}. The solid, curved lines are
separate polynomial fits to the data above and below the empirical
transition temperature of $0.94 T_{\text{o}}$. The measured energy
is actually the kinetic and interaction energies of the trapped
atoms and is measured by suddenly releasing the atoms from the
magnetic trap and calculating the second-moment of the velocity
distribution of the released atoms. (inset) The difference
$\Delta$ between the data and the classical energy emphasizes the
change in slope of the measured energy-temperature curve near
$0.94 T_{\text{o}}$ (vertical dashed line). Figure taken from ref.
\protect{\cite{Ensher96}}.} \label{energy}
\end{figure}

\begin{figure}
\begin{center}
\includegraphics*[width = 0.75\linewidth]{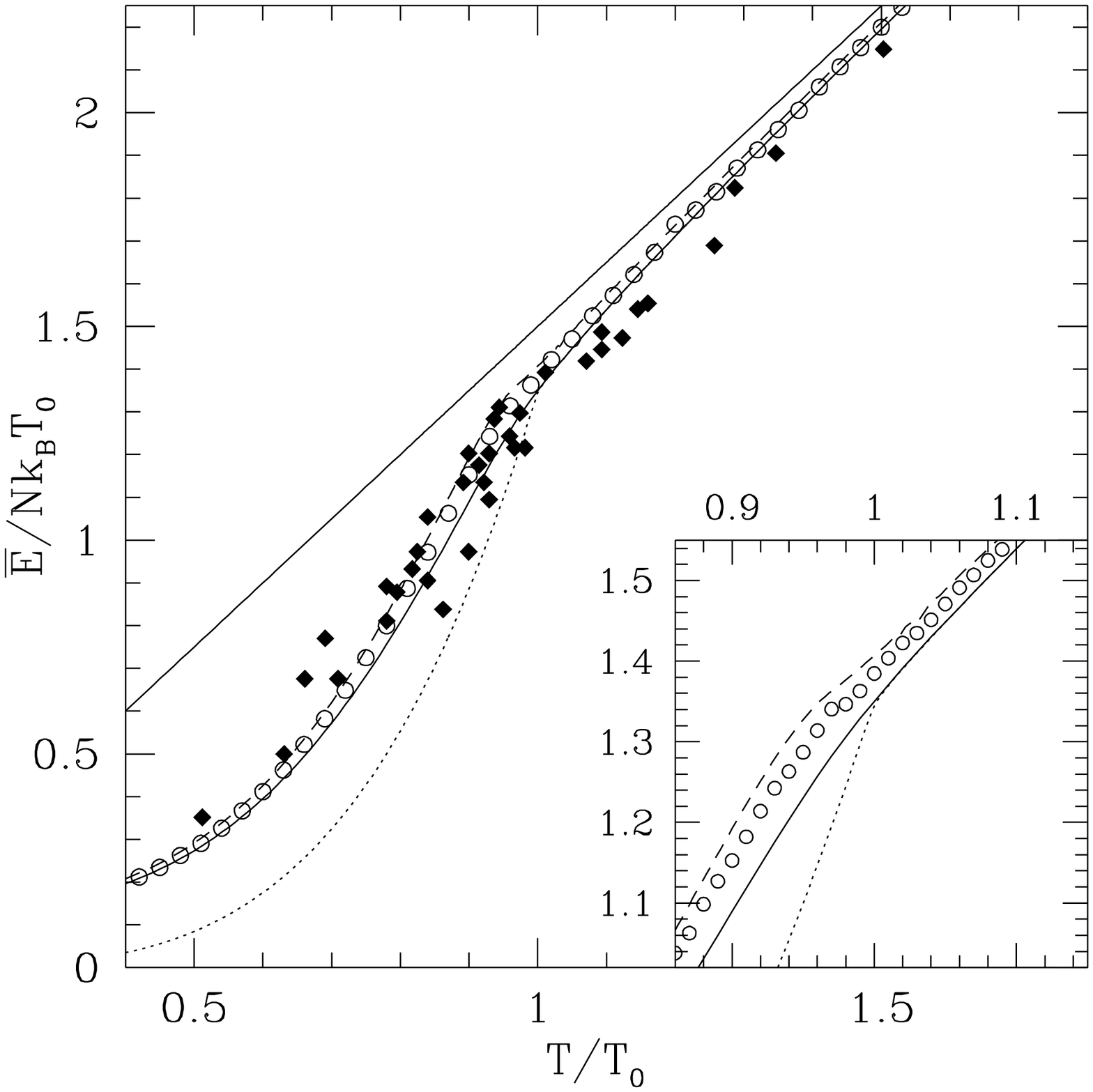}
\end{center}
\caption{Calculation of the kinetic and finite-T interaction
energy of a Bose gas compared with the data of Ensher
\protect{\it{et al} \cite{Ensher96}} (diamonds) and with the ideal
gas result (dotted curve). Results are shown for the zero-order
solution (full curve), the first-order perturbative treatment
(dashed curve) and for the complete numerical solution (circles).
The straight line is the classical Maxwell-Boltzmann result. The
inset is an enlargement of the region around
\protect{$T_{\text{c}}$}. Used with permission of S. Conti from
ref. \protect{\cite{Minguzzi97}}. } \label{energytheory}
\end{figure}

\subsubsection{Energy content}  Interactions will also affect
the specific heat of a confined Bose gas, or equivalently, the
total energy content, which is the temperature integral of the
specific heat \cite{Minguzzi97, Davies97, Giorgini97}. There has
been an experimental study of the energy content as a function of
temperature, based on analysis of images expanded clouds
\cite{Ensher96}. The data (figure \ref{energy}) clearly show a
deviation from ideal gas behavior, which can be accounted for by
finite-$T$ theory (figure \ref{energytheory}). The experimental
procedure is probably not sufficiently accurate to provide a
definitive and quantitative test of the theory.

\subsubsection{Density distributions} Interactions of the
non-condensed fraction with the condensed fraction, and
interactions of the non-condensed fraction with itself, should
modify the spatial distribution of the non-condensed fraction,
particularly just at the edge of the condensate cloud. There exist
some carefully acquired images of this region \cite{Hau98b} but a
theoretical study of the sensitivity of these images to alternate
models of finite-$T$ interactions has not been undertaken.

\subsection{Future directions}
There are a number of exciting experimental paths to explore.

\subsubsection{Finite {$T$}}
As discussed just above, our understanding of finite temperature
interactions would benefit from quantitative measurements taken in
a regime that facilitates comparison with theory.

\subsubsection{Feshbach resonances} \label{Feshbach}
Some years after their significance for BEC was first pointed out
\cite{Tiesinga92, Tiesinga93}, Feshbach resonances have been
observed now in sodium \cite{Inouye98} and in rubidium
\cite{Courteille98, Roberts98}. The interactions between atoms is
now in principle an experimentally tunable quantity.  Whether the
Feshbach resonances will in the end be useful remains an open
question. In sodium at least, there appears to be an enormous
increase in the \emph{inelastic} scattering rate associated with
the Feshbach resonance \cite{Inouye98}. If this turns out to be a
general feature of Feshbach resonances, it will make
tuned-interaction experiments much less convenient.

\subsubsection{Vortices}  Vortices are the heart and soul of
superfluidity.  Experimental studies of vortices will be critical
to our understanding of superfluidity in dilute-gas BEC.

\subsubsection{Many-body effects} Producing experimental evidence
for interactions beyond the mean-field in dilute-gas BEC will be
difficult, but quite worthwhile.

\subsection{A perspective} \label{perspective}
Equation \eqref{gp} above looks superficially similar to the
Landau-Ginzburg equation one finds throughout condensed matter
theory. It is worthwhile to reflect, however, on the important
differences. The coefficient `$a$' in the nonlinear term is not a
temperature-dependent quantity, nor is it a factor empirically
determined from observed fluid behavior.  Rather, `$a$' is the
scattering length for a two-body collision in free space.  For the
species which have to date been made to condense, the value of $a$
has been determined from spectroscopic studies of pair-wise
collisions, entirely independently from the measurements of
collective behavior discussed in this paper.  Equation \eqref{gp},
above, and for that matters Eqns. \eqref{BED}-\eqref{tc}, make
quantitative predictions for Bose condensation behavior with no
adjustable parameters.

The ability to predict macroscopic quantities such as critical
temperatures, healing lengths, specific heats, etc., from
calculations based on independently measured microscopic
interactions, with no adjustable parameters, is a rare occurrence
in condensed matter physics.  Power-law behavior can be extracted
from renormalization group calculations, but the actual values are
almost always determined empirically. It is common for modern
condensed matter physicists to dismiss the actual value of a
critical temperature, or the actual length of a particular
correlation function, as ``uninteresting," when perhaps what they
really mean is ``too difficult."  In any case, we find very
appealing the idea that someone will soon be able to begin with
the elementary two-body collision potential and proceed
systematically through to a quantitative prediction for the
temperature-dependent time constant of the decay of the spiral
motion of a vortex core. The scenario is all the more pleasing
because within a year before or after the prediction is published,
an experimenter will very likely measure that same quantity.

\section{Excitations}

In this section, by ``excitations" we mean ``coherent fluctuations
in the density distribution."  Excitation experiments in
dilute-gas BEC have been motivated by two main considerations.
First, BEC is expected to be a superfluid, and a superfluid is
defined by its dynamical behavior.  Studying excitations is an
obvious first step towards understanding dynamical behavior.
Second, in experimental physics a precision measurement is almost
always a frequency measurement, and the easiest way to study an
effect with precision is to find an observable frequency which is
sensitive to that effect.  In the case of dilute-gas BEC, the
observed frequency of standing-wave excitations in a condensate is
a precise test of our understanding of the effect of interactions.
In this section we will review work to date on excitations,
concentrating almost exclusively on the experimental side.

\subsection{Probing excitations} 
\begin{figure}
\begin{center}
\includegraphics*[width = 0.75\linewidth]{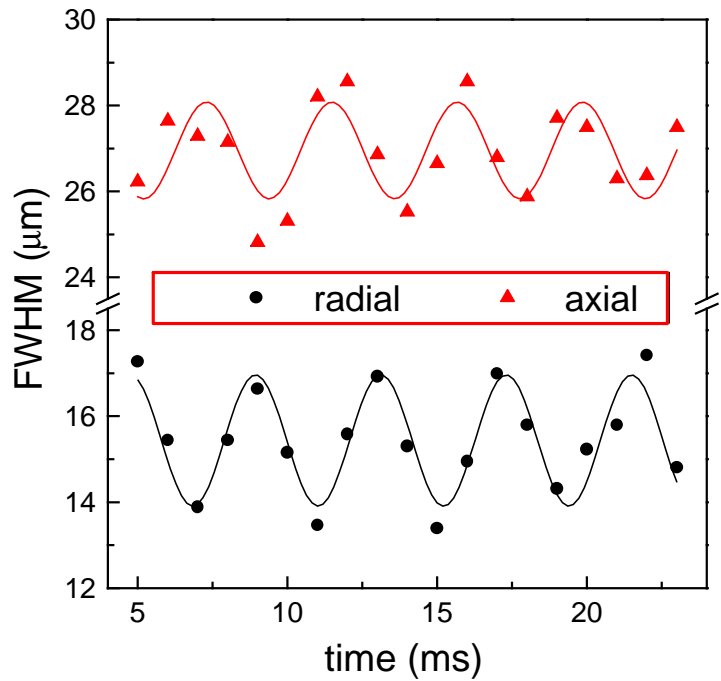}
\end{center}
\caption{Zero temperature excitation data from
\protect\cite{Jin96}. A weak $m=0$ modulation of the magnetic
trapping potential is applied to a $N\approx4500$ condensate in a
132 Hz (radial) trap. Afterward, the freely evolving response of
the condensate shows radial oscillations. Also observed is a
sympathetic response of the axial width, approximately
$180^{\circ}$ out of phase.  The frequency of the excitation is
determined from a sine wave fit to the freely oscillating cloud
widths.} \label{excite1}
\end{figure}

BEC excitations were first
observed in expanded clouds \cite{Jin96}.  The clouds were
coherently excited (see below), then allowed to evolve in the trap
for some particular dwell time, and then rapidly expanded and
imaged via absorption imaging. By repeating the procedure many
times with varying dwell times, the time-evolution of the
condensate density profile can be mapped out. From that data,
frequencies and damping rates can be extracted.  In axially
symmetric traps, excitations can be characterized by their
projection of angular momentum on the axis. The perturbation on
the density distribution caused by the excitation of lowest-lying
$m=0$ and $m=2$ modes can be characterized as simple oscillations
in the condensate's linear dimensions.  Figure \ref{excite1} shows
the widths of an oscillating condensate as a function of
dwell-time.

The use of {\it in situ} imaging is particularly useful in studying
condensate excitations.  In a single measurement, many
observations of the width can be made.  Simultaneous calibration
of the trap frequency can be extracted from observing any residual
dipolar ``slosh" of the condensate (this mode is insensitive to
interactions) superimposed on the driven oscillation. Ketterle's
group has developed this technique to the point where that a
frequency can be determined to much better than $1\%$ in a single
shot \cite{Stamper-Kurn98b}.

Higher-order modes do not result in oscillations of the overall
condensate width but can be observed directly as density waves
propagating through the condensate \cite{Andrews97b}.

\subsection{Driving the excitations}
A frequency-selective method for driving the excitations is to
modulate the trapping potential at the frequency of the excitation
to be excited \cite{Jin96}.  Experimentally this is accomplished
by summing a small ac component onto the current in the trapping
magnets. In a TOP trap, it is convenient enough to independently
modulate the three second-order terms in the transverse potential.
By controlling the relative phase of these modulations, one can
impose $m=0$, $m=2$ or $m=-2$ symmetry on the excitation drive.
The frequency selectivity of this method can in instances be a
disadvantage --- one can miss what one is not looking for. A still
simpler technique for driving the oscillations is to impose a
single ``step-function" on the confining potential
\cite{Mewes96b}.  This approach provides an inherently very
broadband excitation, which has its own advantages and
disadvantages. Condensates can be manipulated in lossless ways by
a focused beam of far-detuned blue light \cite{Davis95b}.  This
technology has been exploited to excite high-order modes in
condensate \cite{Andrews97b}.

\begin{figure}
\begin{center}
\includegraphics*[width = 0.85\linewidth]{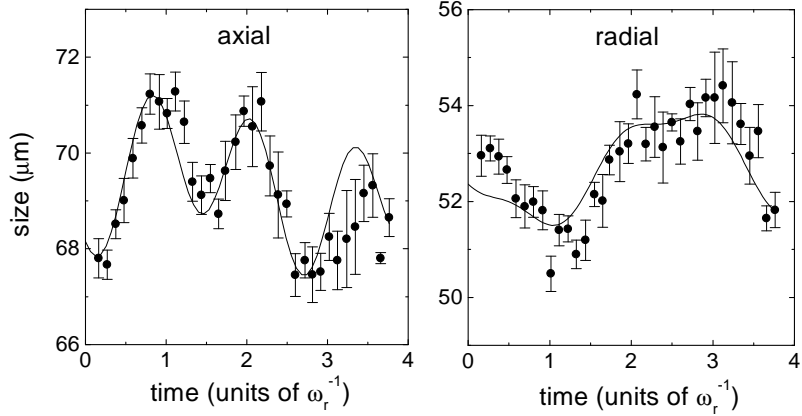}
\end{center}
\caption{Oscillation in the width of a pure BEC cloud in both
axial and radial direction due to the instantaneous change in
scattering length. The widths are for condensates as a function of
free evolution time in units of $\omega_{\text{r}}^{-1}=9.4$ ms,
followed by 22 ms of ballistic expansion. Each point is the
average of approximately 10 measurements. The solid line is a fit
of a Gross-Pitaevskii model to the data, with only the amplitude
of oscillation and initial size as free parameters. Figure taken
from ref. \protect{\cite{Matthews98}}. } \label{excite2}
\end{figure}

The techniques mentioned above all involve exciting the condensate
by manipulating the external potential. Much the same ends can be
achieved by manipulating the internal interactions of the
condensate. One would ideally like to modulate the condensate
interaction strength arbitrarily and at will, for instance by the
use of Feshbach resonances. The JILA group was able to drive
excitations in the condensate by discontinuously transferring the
atoms from one internal state to another \cite{Matthews98}.  The
different internal states had different interaction strengths and
the discontinuous change in mean-field energy resulted in the
simultaneous excitation of two $m=0$ modes. See figure
\ref{excite2}.

\subsection{Connection to theory}
There have been a very large number of theory papers published on
excitations; much of this work is reviewed in \cite{DalfovoXX}.
All the zero-temperature, small amplitude excitation experiments
published to date have been very successfully modeled
theoretically.  Quantitative agreement has been by and large very
good; small discrepancies can  be accounted for by assuming
reasonable experimental imperfections with respect to the $T=0$
and small-amplitude requirements of theory.

\subsection{finite T} \label{finitetempexc}
The excitation measurements discussed above have also been
performed at non-zero temperature \cite{Jin97, Stamper-Kurn98b}.
The frequency of the condensate excitations are observed to depend
on the temperature, and the damping rates show a strong
temperature dependence.  This work is important because it
represents the only observations so far performed that bear on the
finite-temperature physics of interacting condensates. Connection
with theory remains somewhat tentative. The damping rates, which
are observed to be roughly linear in temperature, have been
explained in the context of Landau damping \cite{Liu97,
Fedichev98}. The frequency shifts are difficult to understand, in
large part because the data so far have been collected in an
awkward, intermediate regime: the cloud of noncondensate atoms is
neither so thin as to have completely negligible effect on the
condensate, nor so thick as to be deeply in the ``hydrodynamic
(HD) regime." In this context, ``hydrodynamic regime" means that
the classical mean free path in the thermal cloud is much shorter
than any of its physical dimensions.  In the opposite limit, the
``collisionless regime," there are conceptual difficulties with
describing the observed density fluctuations as ``collective
modes." The MIT Sodium group has published data
\cite{Stamper-Kurn98b} on mixed-cloud excitations which are
tending to but perhaps not safely within the hydrodynamic limit;
the JILA group's published data \cite{Jin97} describe experiments
in which the noncondensate cloud was thin enough to be in the
collisionless limit but which still had distinct effects on the
condensate excitations.  In order to make better contact with
finite-$T$ theory, future experiments could be performed on much
bigger clouds, so as to make the HD limit more accurate.
Alternatively work could be performed in traps with spherical
symmetry \cite{EnsherThesis}, which would make more tractable the
theory of excitations in the collisionless or the intermediate
regime.

\subsection{nonlinear excitations}
The GP equation is manifestly nonlinear; the observations
discussed so far in this section are performed in a small
amplitude, linear limit.  Some data exist characterizing the
lowest-order anharmonic behavior \cite{Jin97, Mewes96b}, but in
general there has been little work on nonlinear dynamics.
Theorists have predicted a range of really interesting behavior.
Varying the aspect ratio of the confining potential causes the
small-amplitude frequencies of various standing-wave modes to move
around. One can arrange for accidental degeneracies -- for
instance at a certain aspect ratio the frequency of the
lowest-order $m=0$ mode is exactly half that of the next
higher-order $m=0$ mode. As one approaches the magic trap
parameter, the anharmonic coefficient diverges and one should be
able to observe frequency doubling effects \cite{Dalfovo97b}. In
the presence of gravity, the aspect ratio of an axially symmetric
TOP trap can be varied from $\sqrt{8}::1$ (oblate) down to
slightly under $1::2$ (prolate) \cite{EnsherThesis}. An IP trap
can be varied from extremely prolate (greater than $30::1$) to
slightly oblate, although in the near-spherical regime, a IP
trap's axial symmetry will be compromised by the effects of
gravity \cite{Monroe90}.

When the condensate is driven very hard, the motion will become
chaotic, and energy will couple into many different modes
\cite{Castin97b}. Dilute-gas BEC may be an instructive model
system for experiments to probe the boundary between chaotic and
truly thermal behavior.

A variety of soliton and soliton-like behavior is also predicted,
and is likely to be observable experimentally. See for example
ref. \cite{Morgan97}.

\subsection{Many-body effects}
As a condensate becomes more dense, the mean-field GP equation
will begin to break down. The first-order corrections to
excitation frequencies have been calculated \cite{Pitaevskii98}.
Precision measurements of condensate excitation frequencies may
provide sensitive tests of this nontrivial theory.

\subsection{Perspective}
The two most recent published experiments have in one case
measured an excitation frequency that agrees with theory to well
under $1\%$ \cite{Stamper-Kurn98b}, and in the other determined
amplitude data from which a ratio of interaction strengths was
extracted \cite{Matthews98}. The ratio agreed with the ratio
determined from two-body spectroscopic data, also to better than
$1\%$.

It is worth pausing to think about what the accuracy of these two
measurements, and the associated theory, tells us about the field
of dilute-gas BEC. As discussed in Section \ref{perspective}
above, this sort of connection between macroscopic interacting
behavior and calculations beginning \emph{ab initio} with two-body
potentials is something a little out of the ordinary.

\section{Condensates and Quantum Phase}

Bose condensates are a wonderful environment in which to
investigate the subtle concept of quantum phase. Experimental work
on  phase is really only just now getting started.  There are
several papers already in print, notably the Ketterle group's
observation of interference between two separate condensates
\cite{Andrews97a} and a collection of papers on mixed-state Rb-87
condensates from the JILA group \cite{ Hall98c}. We recently
reviewed the mixed-condensate work in a separate publication
\cite{Cornell98}, and so will not discuss it here. For the
separate-condensate interference experiment, see the original
paper \cite{Andrews97a} or ref. \cite{Ketterle98}.  We know of
interesting phase experiments going on at Yale \cite{BAnderson98}
and at Gaithersberg \cite{Helmerson98}. Presumably papers from
these groups will have appeared in print by the time this article
is published.

We postpone to a future paper a thorough review of phase
experiments. For the remainder of this section, we would like
instead to discuss some of the terminology of the topic of quantum
phase in condensates, terminology which has led, in our opinion,
to some unfortunate misconceptions. Some may find the content of
this section trivial  --- it is aimed at the confused
nonspecialist.

\subsection{A common misconception} In informal
conversations, we have heard it asserted that a precise
measurement of the number of atoms in a condensate must ``destroy"
a condensate, so that it is ``not a condensate anymore."  The
argument goes like this: (i) the defining characteristic of a
condensate is its coherence; (ii) if you measure the number of
atoms in a condensate very precisely, it will be in a number
state;  (iii) if a collection of particles is in a number state,
it can not be in a coherent state; (iv) since it is not in a
coherent state, it is no longer a condensate! The careful reader
will see the flaw in this syllogism, but in oral discussions,  the
argument can be compelling. The problem of course is that the
phrase ``coherent state" is a technical term from the field of
quantum optics \cite{Glauber63}. It refers to only a small portion
of a broader concept which we could call ``having coherence." It
is certainly true that if the atoms have no coherence, then the
condensate has been destroyed. And it is equally true that a
precision measurement of the number destroys a coherent state. The
fallacy is to equate the destruction of the coherent state with
the destruction of coherence. The fallacy is perpetuated by a
tendency to treat the concept of spontaneous symmetry breaking as
a fundamentalist Truth, rather than as a useful metaphor.

\subsection{The basic phase thought experiment} We turn
now from the topic of what is wrong to the topic of what is
correct.  We will couch our discussion in terms of particular
thought experiments, rather than in the usual language of
expectation values of field operators.  In our experience the
formal notation of the field operators and brackets has sufficient
ambiguity that it as likely to confuse as to enlighten. Here,
then, is the basic experimental module: take a not-too-well
determined number  of atoms from the vicinity of a particular
point in space $(x_1)$, and a similar bunch from from the vicinity
of another point $(x_2)$, and cause cause both bunches to impinge
on an atom counter in such a way as to ensure there is no way the
counter can tell from which point any given atom originated.  The
number of atoms detected at the counter will be sensitive to an
interference term between the amplitude of atoms arriving from
point one and the amplitude of atoms arriving from point two, and
this interference term will be a function of $\text{cos}
\delta\phi (x_1,x_2)$, where $\delta \phi(x_1,x_2)$ is the
difference in local quantum phase between the two points. The
interference term will also depend on the additional differential
phase which may accumulate while the atoms are being transported
to the detector, of course.

\subsection{How is a condensate coherent?} \label{coherentBEC}
In descriptions of condensates, ``coherence" means
``predictability of $\delta \phi$."  A condensate is ``coherent"
because one can predict, for any given time $t_1=t_2$, the $\delta
\phi$ between any two points $x_1$ and $x_2$ within the
condensate.  For a ground-state condensate at rest in the
measurement frame, predicting $\delta \phi$ is particularly easy
because $\delta \phi (x_1,x_2) = 0$ for all $x_1$ and $x_2$ in the condensate.
For a condensate moving with a velocity $v$, one can with
confidence predict that $\delta \phi(x_1,x_2) = m v (x_1 -
x_2)/h$. For a condensate with a vortex, the dependence of $\delta
\phi$ on $x_1$ and $x_2$ is more complicated \cite{vortexphase},
but again completely predictable.  Note that a measure of the
total number of atoms in a condensate, even an arbitrarily precise
measurement, need have no effect on the state of the condensate's
coherence, as defined in this subsection.

\subsection{Relative phase of two condensates}
Although completely irrelevant to the question of whether a
condensate is a condensate, the ``coherent state" of quantum
optics is a useful idea when discussing the \emph{difference in
phase between two different condensates.} In this context, $\delta
\phi$ is indeed the conjugate variable to $\delta N$, the
difference in number between two condensates, and yes, a precise
measurement of $\delta N$ must perforce change unpredictably the
$\delta \phi$ between the two condensates, such that there remains
no coherence between the two condensates. (Although the two
condensates individually can remain ``coherent" objects!)

There is a useful insight due to J. Javanainen
\cite{Javanainen96a} and to Castin and Dalibard \cite{Castinxx}
concerning the relative phase of two condensates. They have
pointed out that, starting with two condensates in a relative
number state, if one proceeds to measure the relative phase, the
experiment will yield a perfectly well-defined precise value.  We
will not review the formalism of their arguments here, but the
gist of it is that the act of sequentially detecting particles
builds up correlations in the remaining particles and
uncertainties in the remaining relative number. After a relatively
small number of particles are detected, the remaining two
condensates find themselves in a relative coherent state.  The
phase of that state is unknowable before the measurement begins:
in the view of Castin and Dalibard, the act of measurement has
projected out, in a quantum mechanical sense, a particular
relative phase.

If we return to the ``building-block" thought experiments on phase
described above, we can see right away what we mean when we say
``a measurement of the relative phase between two condensates
always yields a particular well-defined value." We take two
condensates, called A and B, for simplicity we assume both are
ground-state condensates at rest in the measurement frame, and we
we take a small number of atoms from near a point $x_{\text{A}}$
in condensate A and a small number of atoms from near a point
$x_{\text{B}}$ in condensate B, and beat them on a detector, and
from the interference term, we determine $\text{cos}(\delta \phi
(x_{\text{A}},x_{\text{B}}))$. Immediately thereafter, we pick two
new points, $x_{\text{A}}^\prime$ and $x_{\text{B}}^\prime$, again
in condensates A and B, respectively, and find that we get exactly
the same value for $\text{cos}(\delta \phi)$.  And in general we
find that no matter what point we pick in A and what point we pick
in B, we always find the same relative phase as we found in the
first measurement, although we may not have been able to predict
ahead of time what phase we would measure in the very first
measurement. This then is what we mean by saying that a
``measurement of the relative phase of two condensates always
yields a particular well-defined relative value."  If we assume a
sort of ``transitive property of relative phase," ( {\it i.e.},
that $\delta \phi(x_{\text{A}}^\prime,x_{\text{B}}^\prime) =
\delta \phi(x_{\text{A}}^\prime,x_{\text{A}}) + \delta
\phi(x_{\text{A}},x_{\text{B}}) + \delta
\phi(x_{\text{B}},x_{\text{B}}^\prime)$ ) then the fact that a
measurement of the relative phase of two condensates always yields
a particular, well-defined relative value, is equivalent to the
fact that each condensate is independently ``coherent" in the
sense defined in section \ref{coherentBEC} above.

\subsection{Noncondensate clouds} One can \emph{not} talk about
a single relative phase between two conventional, noncondensed,
thermal atomic clouds (unless one has filtered out all but a
volume within a single thermal coherence length for each cloud --
which in fact is what is done in non-condensate
atom-interferometry experiments.)  The significance of the MIT
two-condensate interference experiment \cite{Andrews97a} and of
other condensate coherence studies \cite{Burt97, Ketterle97} is
that they demonstrated once and for all that condensates are not
simply very cold, relatively dense clouds of atoms.

\subsection{Coherence and decoherence} The relative phase between
two condensates can be determined either by measuring it, as
described above, or by splitting a single condensate into two with
some ``gentle" technology \cite{Matthews98, Hall98c} which does
not too greatly perturb the condensate(s).  Immediately after the
relative phase has been determined, one can of course predict what
will be the result of a new phase measurement.  Because the
relative phase is known, we say that there is a relative coherence
between the condensates. If at some later time we are no longer
able to predict to within an uncertainty of $\pi$ radians the
relative phase, we say the two condensates have undergone
decoherence. The decoherence can arise in many ways. For instance,
if between our initial determination of relative phase and our
final measurement of it, someone should perform a precise
measurement of the relative number, we know the measurement would
cause decoherence.  On the other hand suppose the two condensates
are at slightly different heights, so that they feel slightly
different gravitational potentials.  Their relative phase will
then evolve at a rate proportional to the difference in the
gravitational potential.  If the difference in their heights
remains very constant over time, we can perform several
measurements, determine accurately the rate at which their
relative phase evolves, and thereafter be able to predict at some
future time what their relative phase will be.  If the difference
in their heights varies in time in some uncharacterized way, we
can lose track of the relative phase and be unable to make an
accurate prediction of it.  From an experimenter's point-of-view,
then, decoherence needn't (and usually doesn't) arise from
anything particularly fundamental.

\subsection{Mixtures and superpositions} Experimentally, the
distinction between ``a mixture of two condensates" and a ``single
condensate of atoms in a superposition state,"  is
exactly the distinction between decoherence and coherence.  A
mixture is a superposition in which we have forgotten the relative
phase.  A superposition is a mixture for which we still remember the
relative phase.  The distinction between a mixture and a superposition
has little to do with the intrinsic nature of the sample in question,
and much to do with the state of the experimenter's knowledge.

\subsection{A final warning on vocabulary}
Much has been written about a particular mechanism for
decoherence, one which arises from the nonlinearity of condensate
self-interactions: (i) In a finite-sized condensate, the energy
per atom depends on the number of atoms. (ii) When the phase
between two condensates is measured, an uncertainty in the
relative number of atoms is always introduced. Therefore, there
will be an uncertainty in the difference in the energy per
particle for two systems, and over time an uncertainty in the
relative phase must develop \cite{Leggett95}. The mechanism has
been given the deceptive title ``quantum phase diffusion."  While
the mechanism does involve the quantum phase, and while that phase
does change unpredictably (which is to say, it diffuses),
``quantum phase diffusion" is only one of many ways in which the
phase can become unknowable.  For reasonably sized condensates,
finite temperature effects will almost surely be more important
\cite{Graham98}.

\section{Stray Heating: a Pessimistic note} \label{heating}
\subsection{Introduction to the problem}
Ultracold atoms confined in a magnetic trap have a tendency, in
the absence of ongoing evaporative cooling, to warm up. Given that
the ambient temperature of the trapping environment is typically nine
orders of magnitude higher than the temperature of the trapped atoms, the
surprise perhaps is not that the atoms heat, but that they heat so
slowly.  To an optimist, a heating rate of 100 nanokelvin/sec
means that the time constant for the atoms to thermally
equilibrate with the 300 K environment is about one hundred years.
Hardly objectionable! To a pessimist, (or to a realist) that same
heating rate means that the temperature of a sample initially at
200 nK will double in only two seconds, which can be a major
experimental inconvenience.

The phenomenon of trapped-atom heating is observed, as far as we
know, in all magnetic trapping laboratories. Relatively little has
been published on it, however. In this, the final section of our
paper, we will survey what is known about the various mechanisms
for heating.  While the topic has for the most part not attracted
much scientific study, it is worth investigating for two principle
reasons. First, this ubiquitous effect represents a systematic
which often complicates the interpretation of condensate studies,
such as those on condensate formation or on collisional loss.
Second, heating seems to scale unfavorably with increasing atom
number, and may give rise to an upper limit on the size of
condensates one can produce.

It is difficult to give a satisfactory summary of the experimental
observations. Some quantitative studies are described in
\cite{Myattthesis, Monroethesis}, but most of what is known in the
community has been passed around anecdotally, by word of mouth.
Here are some general observations:  A typical heating rate for a
cloud of $10^7$ Rb-87 atoms in a Ioffe-Pritchard trap, in the
$F=1, m_F=-1$ state, at 1 $\mu$K, is perhaps 150 nK/s. There is
anecdotal evidence, although no careful comparisons, that indicate
the problem is worse in Rb-87 than it is in Na. The problem seems
to be worse in the Rb-87 $F=2, m_F=2$ state than it is in the
$F=1, m_F=-1$ state. Heating is observed to get worse with
increasing trapped-atom number, with increasing density, with
increasing column density, with increasing elastic collision rate,
and with decreasing bias field. Unfortunately, it is difficult to
vary each of the above-mentioned quantities independently,
therefore it is not clear what really matters.  Both the MIT
Sodium group and the JILA BEC group have both observed that the
presence of an ``rf shield" (described in Subsection
\ref{rfshield} below) profoundly affects the heating rate, in most
situations.  Heating rates as low as 10 nK/s have been observed in
well-shielded traps \cite{Stamper-Kurn98b}

Efforts on the part of the JILA group to come up with an
experimentally validated, comprehensive model for the heating
mechanism(s) have not been very successful. Our experimental
results have been somewhat inconclusive, and our models tend to
become too complicated to give simple numerical predictions.  The
discussion in this section is meant to be taken as a collection of
general observations, as suggestions for directions of research,
rather than as definitive results.

In the remainder of this section we discuss in turn various
candidate mechanisms for heating. We believe most of these
mechanisms are inadequate by themselves to account for the
observed heating. In the final subsection we explore what we feel
is the best explanation for the heating --- the ultra-cold trapped
sample is often surrounded by a very dilute haze of much hotter
atoms, also trapped, which gradually thermalizes with the
ultra-cold sample. Glancing collisions with background atoms, and
inelastic decay products, while inadequate to directly cause
heating, help populate the high-energy haze and thus indirectly
contribute significantly to the heating.

\subsection{Miscellaneous small effects}
\subsubsection{Anti-evaporation} The earliest paper on trapped-atom
evaporation \cite{Hess86} discusses one of the most ubiquitous
mechanisms for heating: inelastic collisions. These can be thought
of as giving rise to ``anti-evaporation" for the following reason:
at the typical temperatures of trapped atoms, the rates for
three-body recombination and dipolar relxation are independent of
the collision velocities, and the per-volume rates are simple
functions of the local density.  Since in thermal equilibrium the
density is highest where the potential is deepest, inelastic
collisions selectively remove atoms with low potential energy,
thus increasing the mean energy of the remaining atoms.  Heating
from anti-evaporation is very common; most experimenters have
probably encountered it.

In samples below the critical temperature, most of the inelastic
collisions occur in the condensate, because its density is much
higher than the thermal cloud. If the sample is not too large, the
decay products can pass cleanly out of the cloud.  In this case,
the inelastic collisions consume the condensate but do not
strictly speaking cause heating, because the density and
temperature of the thermal cloud is unchanged.  But because the
overall number of atoms is decreasing, the value of the critical
temperature goes down.  Antievaporation therefore increases the
value of $T/T_{\text{c}}$.

Antievaporation usually arises from inelastic collisions, but more
generally anything which preferentially removes atoms from the
center of the trap gives rise to antievaporation.  For instance,
if there is trap loss due to rf magnetic noise driving unwanted
Zeeman transitions, and if that noise has a steep frequency
dependence (say, due to low-pass filters on the trap-magnet
coils), then atoms at the lower magnetic fields at trap center
will undergo Zeeman transitions and leave the trap at a faster
rate.

In many experimental situations, heating is observed that is far
too large to be accounted for by antievaporation alone. In
anti-evaporation's limiting case, only atoms with zero mean energy
would undergo decay. A differential change of number $\delta N$
would then give rise to a differential increase in temperature
$\delta T$ such that $\delta T/T = -\delta N/N$.  This is an
\emph{upper limit} on the amount of heating that can arise from
anti-evaporation.  When the observed fractional rate of heating is
larger than the observed rate of trap loss, one must look
elsewhere to identify the dominant source of heating.

\subsubsection{Field noise/Shaking trap}
Atoms in a magnetic trap are by intention highly isolated from the
external environment. The one mechanical force that {\em must} be
present is from the trapping itself, and thus in investigating
heating we naturally suspect the trapping fields. In our
experience, however, mechanical noise on the magnetic coils, or
electric noise in the coils, is seldom the dominant source of
unwanted heating. There are two reasons for this -- (i) moving the
coils, or modulating the current in them, affects all the atoms
nearly the same, and thus does not couple well to a heating
effect, except to the extent that it drives macroscopic pulsing or
sloshing modes of the cloud. (ii) In magnetic traps at least, and
at low temperatures, the atoms see a very harmonic potential. This
has the effect of sharply constricting the bandwidth of the noise
that can harm the atoms. Unless the noise is near a trap
oscillation frequency, or a harmonic thereof, the energy can't
couple to the atoms.

If the noise is at the second harmonic of trap frequency, the
motion of the atoms can be driven parametrically, and the
resulting ``pulsing" of the cloud can convert via collisions to
heat. If the noise is directly at a trap frequency, ``sloshing"
can be driven, which due to anharmonicity also converts to heat.
Directly driven sloshing is usually only a problem in the vertical
direction, for which gravity breaks symmetry. In an effort to
characterize the heating from noise-driven pulsing, we have
intentionally applied electrical and mechanical noise to coils. If
the noise is broad-band, it takes a surprisingly high power
density to accomplish any heating.

One week at JILA in 1997, when we were having a particularly bad
problem with electrical ground loops, our evaporation stopped
working well. It turned out that during our evaporation procedure
(which includes an adiabatic ramp of the trapping frequencies) we
were unintentionally pausing for a second at the value of the trap
parameters which caused one of the trap frequencies to be resonant
with a multiple of 60 Hz (the North American power-line
frequency).  Eventually we fixed the power supply, but we were
able to implement a successful ``quick fix" simply by ramping very
quickly over the offending region of trap parameters. In nine
years of JILA experience with magnetic traps, this was the only
occasion in which magnetic coil noise was confirmed to be the
dominant source of unwanted heating.  We have used parametric
driving to deliberately heat the atoms (see {\it e.g.} ref.
\cite{Monroe93}), but to accomplish this we had to apply a
modulation tone of considerable amplitude compared to background
noise.

For a thorough, quantitative discussion of heating from noise on
the trapping field, see Savard \emph{et al.} \cite{Savard97}.
While they analyze the same mechanisms we mention above, they come
to the opposite qualitative conclusion --- they assert that noise
on a trap is likely to be an important source of heating.  The
source of the disagreement lies in the technical details  -- their
model system was an optical rather than a magnetic trap, and
optical traps tend to have higher confining frequencies. As Savard
\emph{et al.} show, the higher the confining frequency, the more
susceptible a system is to parametric heating \cite{Savard97}.
Moreover, electrical and mechanical stability of electromagnets is
likely to be better than intensity and pointing stability in laser
beams.

Noise-induced heating is likely to be a very important effect in
``composite traps," traps made of magnetic and optical forces
\cite{Davis95b} or of more than one laser beam. In those cases
noise can lead to relative motion of the various components of
the confining potential, and thus heat the atoms more directly.

\subsubsection{Stray rf and optical fields}
If care is not taken to keep radio and optical frequency photons
out from where they do not belong, they can cause trap loss and
presumably, under the right conditions, heating as well. In
practice it is often necessary to place an opaque box around
either the lasers (with their associated absorption cells) or the
magnetic trap.

\subsubsection{Black-body radiation}
In conventional cryogenic experiments, black-body radiation is a
major source of heat transfer, but it is difficult to imagine that
it will ever be important for the case of trapped atoms. The
cross-section for scattering a photon, at least for atoms, is
simply too small.  The question may need to be revisited if one
were to attempt to evaporatively cool objects more complicated
than individual atoms, objects which may couple more strongly to
long-wavelength radiation.

\subsection{Collisions with background atoms} \label{bgcollisions}
Another common source of unwanted heating in conventional
cryogenic experiments is imperfect vacuum: residual gas atoms move
back and forth between warmer and colder surfaces, transferring
heat with each bounce.  In trapped-atoms experiments as well the
room-temperature residual gas atoms can provide heat to the
confined atoms, but the effect can not be understood as a simple
shuttling of heat between a warm surface and a cold sample. When a
300 Kelvin background atom collides with a trapped atom, the most
probable result will be a clean ejection -- both the incoming and
the impacted atoms leave the trap without further collisions.
These are the events that give rise to the background loss rate
$\gamma_{\text{bg}}$ discussed in section \ref{collisionscaling}
below. But there exists also the possibility of glancing
collisions which transfer a relatively small amount of energy to
the trapped atom. These events lead to heating rather than to
loss.

\subsubsection{Direct creation of ``hot" atoms} The distinction between
heating and loss can be somewhat arbitrary. Imagine a magnetic
trap which is 5 mK deep, confining at its center a sample of atoms
with temperature 20 $\mu$K. Now suppose a trapped atom undergoes a
collision that leaves it with 4 mK of kinetic energy. Is this
heating, or loss?  If one were to measure the total number and
total energy of the atoms confined in the potential, this
particular collision looks like heating. A more common
experimental procedure, however, would be to image the cloud, and
fit its profile to a Gaussian lineshape. If one identifies the
energy of the cloud as the mean-square width of the fit lineshape,
and the number of atoms in the cloud as the area under the
lineshape, then our hypothetical collision looks like loss.
Typically, then, ``heating" results from background collisions
which transfer energy less than three or four times the mean
energy of the trapped cloud.  At the opposite extreme, collisions
which leave the impacted atom with energy larger than the trap
depth clearly result in loss.  Collisions which transfer an
intermediate range of energy give rise to a large-area, diffuse
cloud of atoms that are still trapped but whose optical density is
below detection threshold. These intermediate collisions appear to
give rise to loss, rather than heating, at least at short times.
The longer-term effects of the diffuse cloud are discussed in
subsection \ref{Oortparadigm} below.

To get a rough estimate of the size of these heating effects, we
need a simple model for small-angle collisions with background
gas. The following simplified explanation draws on thorough
treatments by Helbing and Pauly \cite{Helbing64} and by Anderson
\cite{Anderson74}; our results are not meant to be accurate to
within a factor of two. Small-angle collisions are the result of
trajectories with large impact parameters. Therefore only the
long-range tail of the inter-atom potential is relevant. We write
the inter-atomic potential as $U_{\text{int}}=C_6/r^6$. The value
of $C_6$ depends on the scattering species. For helium on
rubidium, $C_6$ is 36 atomic units (au) \cite{Kleinekathofer96}.
For rubidium on rubidium, the value is 4700 au. Treating the
scattering event classically, in the impact approximation, and
working in the lab frame (in which the trapped atom is initially
at rest, and the incoming background atom has energy
$E_{\text{col}}$), we find the partial cross section for a
glancing collision to transfer energy $E_{\text{t}}$

\begin{equation}
    \sigma_{\text{g}}^{\text{class}}(E_{\text{t}})=\frac{\pi}{6}\biggl(\frac{9{C_6}^2}{E_{\text{col}}}\biggr)^{1/6}
    E_{\text{t}}^{-7/6}.
\end{equation}

\noindent Note that the integral of
$\sigma_{\text{g}}^{\text{class}}(E_{\text{t}})\,\text{d}E_{\text{t}}$
diverges at low energy; quantum mechanics intervenes to keep the
true total cross-section finite.  We define a cross-over energy
$E_{\text{c}}$: for $E_{\text{t}} < E_{\text{c}}$, the deBroglie
wavelength of the transferred momentum becomes comparable to or
longer than the classically determined impact parameter, and
diffraction effects dominate. $E_{\text{c}}$ is given by

\begin{equation}
    E_{\text{c}}\approx2\frac{\hbar^{12/5}}{C_6^{2/5}}\frac{1}{m\,m_{\text{b}}^{1/5}}E_{\text{col}}^{1/5}
\end{equation}

\noindent where $m$ is the mass of the trapped atom,
$m_{\text{b}}$ is the mass of the background atom, and it is
assumed that $m_{\text{b}}\leq m$. For helium on rubidium,
$E_{\text{c}}$ is about 40 mK. For rubidium on rubidium it is
about 3 mK.

As discussed in ref. \cite{Anderson74}, for
$E_{\text{t}}<E_{\text{c}}$, $\sigma_{\text{g}}(E_{\text{t}})$
shows oscillatory behavior due to diffraction.  In our simplified
model, we use the following approximation:

\begin{equation} \label{crosssec}
    \sigma_{\text{g}}(E_{\text{t}})=
    \begin{cases}
        \alpha      &\text{if $E_{\text{t}}\leq E_{\text{c}}$;} \\
        \alpha\bigl(\frac{E_{\text{c}}}{E_{\text{t}}}\bigr)^{7/6} &\text{if
        $E_{\text{t}}>E_{\text{c}}$;}
    \end{cases}
\end{equation}

\noindent where $\alpha =
\sigma_{\text{g}}^{\text{class}}(E_{\text{c}})$. To convert the
cross-section into a rate, one needs to know the density of the
residual 300 K atoms. In the lab, it is easier to measure the
lifetime of a trapped sample than it is to get an accurate
estimate of the local residual vapor pressure at the trap.  If we
define $\gamma_{\text{bg}}$ to be the loss rate measured in the
limiting case of a very low-density sample and a very shallow
trap, we know that $\gamma_{\text{bg}}$ will be proportional to
the integrated collision rate and to the background pressure.
Assuming there is one dominant species of background gas, we can
normalize eqs. \ref{crosssec} to get the following {\em partial
rate} results, for the rate per atom, per differential energy, of
energy transfer collisions:

\begin{equation} \label{etcollrate}
    \gamma_{\text{g}}(E_{\text{t}})=
    \begin{cases}
        \frac{\gamma_{\text{bg}}}{7 E_{\text{c}}} &\text{if $E_{\text{t}}\leq E_{\text{c}}$;} \\
        \frac{\gamma_{\text{bg}}}{7 E_{\text{c}}}\bigl(\frac{E_{\text{c}}}{E_{\text{t}}}\bigr)^{7/6}
        &\text{if $E_{\text{t}}>E_{\text{c}}$.}
    \end{cases}
\end{equation}

\noindent What then is the resulting heating rate from such
collisions? As discussed above, depending on exactly how one
measures ``heating," one can estimate a threshold value, call it
$E_{\text{h}}$, for the transfer energy. For $E_{\text{t}} >
E_{\text{h}}$, the collisions result in ``loss." For
$E_{\text{t}}<E_{\text{h}}$ collisions result in ``heating." The
fractional rate of heating is then

\begin{equation}\label{fracheating}
    \frac{1}{\overline{E}}\frac{\text{d}\overline{E}}{\text{d}t}=\frac{1}{\overline{E}}\int_0^{E_{\text{h}}}
    {E_{\text{t}}\gamma_{\text{g}}(E_{\text{t}})}\,\text{d}E_{\text{t}}
\end{equation}

\noindent where $\overline{E}$ is the mean energy per atom in the
trapped sample. If we take a plausible estimate for the
$E_{\text{h}}$, say $E_{\text{h}}\approx 3\overline{E}$, then
$(\text{d}\overline{E}/\text{d}t)/\overline{E} \approx 2
\gamma_{\text{bg}}kT/E_c$, for a sample with temperature $T$.

This simple model of direct heating from glancing collisions is
qualitatively successful in accounting for heating observed in
early JILA experiments on trapped Cs \cite{Monroe93,
Monroethesis}.  By heating the vacuum chamber, we caused the
dominant background species to be cesium. In this limit, we
observed that the fraction heating rate of a 200 microKelvin cloud
scaled linearly with the background loss rate, with a constant of
proportionality of about 0.5. For cesium on cesium, $E_c$ is about
1 mK, and the predicted constant of proportionality is between
heating and loss is about 0.2. Given the qualitative nature of the
model, the agreement is pretty good. When we instead cooled the
vacuum chamber the dominant background species became helium, and
(consistent with the large value of $E_{\text{c}}$ for helium on
cesium) the heating rate essentially vanished, even though we
continued to observe trap loss.

In most modern BEC experiments, however, the lab conditions are
quite dissimilar to the early JILA Cs experiments, and the simple
model is empirically found to be completely inadequate.  In a
clean vacuum chamber, the dominant background species is usually a
low-mass, low-$C_6$ species such as helium or hydrogen, such that
the appropriate value for $E_{\text{c}}$ may be 40 mK or higher.
The observed value of $\gamma_{\text{bg}}$ is typically less than
0.01 sec$^{-1}$. In addition, the sample temperature is typically
1 $\mu$K or lower. Under these conditions, eq. \ref{fracheating}
typically predicts heating rates four orders of magnitude smaller
than what is observed.

\subsubsection{Heating from secondary scattering.}
As a first attempt to improve the model, we revisit its implicit
assumption that collision products can pass unimpeded through the
sample, {\em i.e.} that the sample size $l$ is much less than the
mean-free-path $\lambda_{\text{mfp}}$.  If the opposite limit
applied, if $l\gg\lambda_{\text{mfp}}$, then an atom which has
acquired an energy $E_{\text{t}}$ from a background event would
collide with other atoms in the sample many times along its
trajectory out of the sample. If the sample is thick enough, the
entire energy $E_{\text{t}}$ could be distributed among a shower
of atoms, all with energies less than $3 \overline{E}$.  We can
get a rough estimate of the resulting heating rate in this limit
from eq. \ref{fracheating}, with $E_{\text{h}}$ now defined as the
maximum value of energy an atom in the middle of the sample can
have and still be trapped by multiple collisions in the dense
sample. $E_{\text{h}}$ is now a function of the column density of
the sample, $nl$, and to estimate its value we look at the energy
dependence of
$\lambda_{\text{mfp}}=(n\sigma_{\text{hc}}(E))^{-1}$, where
$\sigma_{\text{hc}}(E)$ is the energy-dependent cross-section for
large-angle scattering. In the limit of very large $nl$,
$E_{\text{h}}$ will be sufficiently large that
$\sigma_{\text{hc}}(E_{\text{h}})$ will be in the classical limit,
for which $\sigma_{\text{hc}}(E)\varpropto C_6^{1/3}E^{-1/3}$.
With each collision, the particle loses perhaps $1/2$ its energy,
so that $\sigma_{\text{hc}}(E)$ becomes larger and the
corresponding value of $\lambda_{\text{mfp}}$ becomes shorter. The
sum over the sequence of successively shorter values of
$\lambda_{\text{mfp}}$ converges, such that most of the shower of
atoms resulting from the original energy-transfer event should be
brought to rest within a distance proportional to
$E_{\text{t}}^{1/3}/n$. Therefore the maximum energy atom that can
be trapped, $E_{\text{h}}$ scales as $(nl)^3$. Substituting this
value for $E_{\text{h}}$ into eq. \ref{fracheating}, we see that
in the limit of a trapped cloud with very high column density,
$nl$, the heating rate can scale as $(nl)^6$.

A heating rate that scales as the sixth power of the column
density is an alarming prospect, but when reasonable numerical
values are plugged in, we find that our multiple-scattering model
predicts dramatic effects only for very large trapped samples --
with column-densities of perhaps $10^{13}$ atoms/cm$^2$ or
greater. We include the above discussion only because it
illustrates how, even with perfect rf-shielding (see subsection
\ref{rfshield} below), heating may eventually prove to be a
formidable upper limit to the sizes of condensates one can create.
We wish to reemphasize that the models above are very crude in
nature. We made a number of simplifying assumptions which we did
not explicitly state.  For instance, an implicit assumption in the
previous paragraph is that, after a particular background scattering
event, either all the energy will be captured in the sample by
multiple rescattering, or all the energy will escape. In actual
fact, most scattering events will leave behind as heat some
fraction of their initially transferred energy, a fraction that is
neither zero nor unity. Such events are not properly accounted for
in our model.

To conclude this subsection, we should emphasize that for the
column-densities of the trapped samples in currently ongoing BEC
experiments, the amount of heating that can be attributed to the
explicit scenario described above (a {300 K} atom scatters from a
trapped atom, which in turn rescatters on its way out of the
sample) is relatively small.  On the other hand, while it is true
that most of the atoms which are perturbed by background gas will
exit the {\em sample} without additional scattering, a significant
fraction of them may remain within the much larger volume of the
{\em magnetic trap}. These atoms may cause trouble later on, as we
discuss in subsection \ref{Oortparadigm} below.

\subsection{Collisions with the products of inelastic decay.}
\label{inelasticcollisions}

When at atom is lost from the sample due to a collision with a
{300 K} background atom, the chances are $6/7$ that it will have
an energy larger than $E_{\text{c}}\approx40$ mK (see eq.
\ref{etcollrate}). This means, in turn, that the large majority of
atoms lost due to background collisions will have energies greater
than the depth of the confining magnetic trap.  With inelastic
decay, however, the story is reversed. If we assume most
three-body decay populates the least-bound vibrational state, then
the decay products will have energies mostly under 1 mK. Dipolar
decay products may also have energy under 1 mK. Inelastic loss
then, leads to production of atoms with energies typically much
greater than the trapped sample, but much less than the depth of
the confining magnetic potential.  Given that lifetimes due to
background decay often exceed 100 seconds, and given that the
three-body decay constants for alkalis are in the vicinity of
$10^{-29}$, it is often the case that, as the density of a trapped
cloud approaches the critical value for BEC, the dominant loss
process is inelastic decay, rather than background collisions. The
combined result then is that as a sample approaches the critical
density, the dominant mechanism for the production of atoms with
``dangerous" energies ({\em i.e.} $1\mu \text{K}<E<2 \text{mK}$)
will be inelastic decay, rather than glancing collisions with
background atoms.

The decay products will help populate a diffuse cloud of hot
atoms, which can cause heating, as discussed below.  Depending on
the column-density of the trapped sample, they may also contribute
to direct heating, exactly as background-scattered atoms can.
There is a widely-held assumption that the decay products of
three-body recombination are so energetic that they have a
relatively small cross-section for colliding with a trapped atom.
This is completely incorrect. Given that the dominant decay
channel is into the highest vibrational state, a simple
calculation for Rb-87 shows that the outgoing ``witness" atom will
have a cross-section for scattering which is about the same as the
zero-energy s-wave cross-section \cite{Burkecomm}. Beyond the
specifics of the rubidium potential, general considerations on the
relationship between s-wave scattering and the binding energy of
the highest vibrational level suggest that the witness atom will
usually have an energy low enough that its cross-section will be
comparable to the zero-energy value. As for the outgoing dimer,
the kinetics of three-body recombination is such that the dimer
will have per-atom translational energy four times smaller than
the witness atom's energy. There is no reason to believe that the
dimer's cross-section on its atomic counterpart will be small.
After all, the loosely bound dimer is presumably no less
polarizable than a single atom.

\subsection{The Oort cloud paradigm} \label{Oortparadigm}
The usable volume of a Ioffe-Pritchard magnetic trap can be quite
significant, perhaps 2 cm$^3$ or more, and the depth of the
confining potential is typically on the order of several
millikelvin. A typical sample of ultra-cold atoms, with
temperature on the order of a microkelvin, occupies only a tiny
fraction of the available volume: the ratio of sample volume to
the trap volume may be $10^{-5}$ or less. There exists
considerable evidence that, towards the end of an evaporative
cooling cycle when the central sample is nearing degeneracy, the
outer region of the trap is filled with a very dilute, very
high-energy halo of trapped atoms which we refer to as the ``Oort
cloud" \cite{Oortref}. The millikelvin Oort cloud is of course not
in thermal equilibrium with the microkelvin ultra-cold sample, and
the coupling between them is weak, due to the Oort cloud's very
low density. Nonetheless occasional collisions may transfer energy
(and atoms) between the two trapped components, and we believe
this coupling provides the dominant mechanism for sample heating
in many experimental situations \cite{Oortexists}.

The Oort cloud is not readily imaged. Its radius may be ten to 100
times larger than that of the ultracold sample, which means that
its cross-sectional area may be 100 to $10^4$ times larger.
Thus even if the Oort cloud comprises more atoms than the
ultracold sample, its optical depth may be so small as to make it
easy to overlook. In our experience, a thermal cloud that is
allowed to remain in a magnetic trap for several seconds in the
absence of evaporative cooling will sometimes develop ``wings,"
which can not be fit with a single Gaussian density profile. While
these wings are usually thought of as a high-energy tail to the
ultracold distribution, they may as well be described as a
low-energy tail to the Oort cloud.

Direct evidence for the existence of the Oort cloud can be found
in discrepancies between two distinct methods of measuring the
number of trapped atoms in a magnetic trap. The first method is to
pass a probe beam through the trap and image the two-dimensional
absorption profile. The number of atoms in the trap is then
assumed to be proportional to the integrated signal under the
central feature in the shadow cast by the ultra-cold sample. This
method is not sensitive to the presence of an additional,
spatially dispersed population of atoms. The second method is
suddenly to turn off the magnetic trap and to turn on the MOT, and
then to collect the total fluorescence emitted from the atoms
recaptured in the optical trap. The MOT laser beams are several cm
in diameter and sweep up any atoms that had been trapped anywhere
within the magnetic trapping volume.  The two methods can be
performed on similar clouds of trapped atoms, and the results
compared. What is often found is that when a cloud of trapped
atoms has just been freshly loaded into the magnetic trap, the two
methods give similar estimates of the total number of trapped
atoms. Later, after extensive evaporation has been performed and
the number of atoms in the sample reduced 100-fold, the
MOT-recapture method may show perhaps five times more atoms than
the shadow-image method. In this example, then, only 20\% of the
trapped atoms are in the ultracold sample. The remainder populate
the Oort cloud.

\label{rfshield} The strongest evidence that the Oort cloud is
responsible for much of the typically observed heating comes from
the effect of an ``rf shield."  Both at JILA \cite{Myattthesis,
Burt97, EnsherThesis} and at MIT \cite{Mewes96a, Stamper-Kurn98a}
it has been noted that the application of an rf field can affect
the lifetime and heating rate of trapped samples.  During
evaporation, an rf magnetic field is applied to the trapping
region, and its frequency is gradually ramped downward as
evaporation proceeds.  There is a 2-dimensional surface
surrounding the center of the trap, a surface over which the
magnetic field has the correct magnitude to bring the atoms on
resonance with the rf-field. Atoms whose trajectories cross this
surface have some probability of being transferred into an
untrapped spin state. As the frequency of the rf is decreased, the
radius of the rf-surface decreases, cutting into the edge of the
ultracold sample, and driving the evaporative process.  In a
typical experiment, the rf surface is smoothly brought inward
until the ultracold sample reaches the desired temperature, then
moved back out again and left on while experiments are performed
on the sample. This post-evaporation configuration is referred to
as an ``rf shield."  The radius of the rf surface is typically
sufficiently large that the rate of evaporation essentially
vanishes; no ultracold atoms have trajectories with sufficiently
high energy to reach the rf shield.  Yet the shield is observed to
have a profound effect -- turn the rf field off, and the rates of
heating and loss from the ultracold sample can increase by an
order of magnitude.  Presumably the rf shield depletes the Oort
cloud or prevents it from being populated in the first place.

How is the Oort cloud populated? We speculate that there may be
several mechanisms. First, there is the mechanism of glancing
collisions. This is discussed in section \ref{bgcollisions}.
above. Second, there is the generation of inelastic decay
products. This is described in Section \ref{inelasticcollisions},
above. Third,  ``incomplete evaporation."  In experiments with
atoms in the $F=2, m_F=2$ state, rf evaporation may remove an atom
from the $m_F=2$ state but fail to transfer it into an untrapped
state, leaving it instead in the $m_F=1$ state. Some of these
atoms can later cause trouble. Fourth,  ``primordial remnants." At
any given instant, a small fraction of atoms in an optical
molasses have energies which are extremely superthermal. These
atoms may end up loaded into the magnetic trap with trajectories
that are outside the initial radius of the evaporative rf knife.
While these atoms may initially represent less than 1\% of the
total sample, the process of evaporation reduces the size of the
trapped sample by a factor of 100, which increases proportionally
the relative significance of the high-energy remnants. Fifth, the
Oort cloud can populate itself through erosion of the ultracold
sample. For example, an Oort atom with 500 $\mu$K of energy can
plunge into the ultracold sample and undergo a collision, with the
result that two, 250 $\mu$K atoms emerge.  The net population of
the Oort cloud is increased, and it is left colder and denser. The
colder is the Oort cloud, the more tightly is it coupled to
ultracold sample, and the more heating it will cause.

How is the Oort cloud depopulated?  The presence of the rf shield
(or equivalently, ongoing rf evaporation, which amounts to the
limiting case of a close-in rf shield) will tend to deplete the
Oort cloud.  As the atoms pass through the surface in space on
which the rf is resonant, they are transferred into untrapped
states, and may leave the trap. In a TOP trap, atoms in the Oort
cloud may encounter as well the orbiting zero-field point, which
can induce Majorana transitions \cite{Majorana31} and detrap
atoms.

The rf shield is an imperfect device for removing the Oort cloud.
If the amplitude of the rf is too low, then the Oort atoms (whose
average velocity is high compared to the atoms undergoing
evaporation) may pass through the rf's resonant surface too
rapidly to undergo a transition. The trajectories of such Oort
atoms may carry them through the ultra-cold cloud many times
before they are finally ejected by the weak rf field. On the other
hand, if the amplitude of the rf is too high, such that there is
unity probability for an Oort atom to undergo a transition as it
passes through the surface, the situation is equally
unsatisfactory.  As the Oort atom passes through the resonant
surface, towards the center of the trap, it is flipped into an
untrapped state. But its energy will be such that it can still
reach the ultracold atoms at the center of the trap. To leave the
trap, it must pass once again through the resonant surface, where
with high probability it will be flipped back into a trapped
spin-state! See fig. 5b of ref. \cite{vanDruten96} and
accompanying text for a nice description of the relevant physics.

A quantitative understanding of the interplay between the rf
shield and the Oort cloud, and of the consequences for trap loss
and heating, may require a fairly elaborate model.  Certainly any
attempt to create a condensate with say $10^9$ Rubidium atoms will
need to consider the problem of heating very carefully. Perhaps
optical dipole traps, which are so shallow that
they can not support an Oort cloud \cite{Stamper-Kurn98a}, may
play an important role in the creation of extremely large
condensates.

\section{Collisions and Evaporation: an Optimistic note}
The crucial issue in planning a successful dilute-gas BEC
experiment is collisions. Until a new technology comes along, the
last step in a BEC experiment will be evaporation, and evaporation
works only if the collisional properties of one's system are
favorable.  In practice, this means that planning a BEC experiment
requires learning to cope with ignorance. So much is currently
known about the elastic and inelastic collisional properites of
the atoms in the first column of the periodic table
\cite{Heinzen98} that it is easy to forget that essentially
nothing is known about the low-temperature collisional properties
of any other atomic or molecular species.

One can not expect theorists to relieve one's ignorance --
inter-atomic potentials derived from room-temperature spectroscopy
are not adequate to allow theoretical calculations of cold elastic
and inelastic collision rates, even at the order-of-magnitude
level.  Cold-atom spectroscopy can do the job (as we have seen
with lithium, sodium, and rubidium), but starting an experimental
program to investigate the cold collisional properties of a new
atom is a major and uncertain endeavor.  In most cases, the
easiest way to discover if evaporation will work for a particular
atom or molecule is simply to try it.

Launching a major new project without any assurances of success is
a daunting prospect, but we believe that, if one works hard
enough, the probability that any given species can be
evaporatively cooled to the point of BEC is actually quite high.
In this section we will review the scaling laws behind this
optimistic assertion.

\subsection{Collisional scaling predicts success!}\label{collisionscaling}
\label{evapcriteria} The extensive literature on evaporation will
not be reviewed here. See refs. \cite{Hess86, Doyle91} for
important early work. The results of the earliest Monte Carlo
trajectory simulations are presented in refs. \cite{Monroe93,
Monroethesis}. See ref. \cite{vanDruten96} for a useful review
paper. The literature can be summarized in the following
requirement: In order for evaporative cooling to succeed, there
must be a great many elastic collisions per atom in trap per
lifetime of the atoms in the trap. Exactly what is meant by a
``great many" depends on how evaporation is implemented, but a
reasonable guess is ``$200$," give or take a factor of four. Since
the lifetime of the atoms in the trap is usually limited by
collisions \cite{lifetimenote}, the requirement can be restated:
the rate of elastic collisions must be about two orders of
magnitude higher than the rate of bad collisions.

At sufficiently low collision energy, the cross-section for
elastic collisions becomes energy-independent, and the rate per
atom can be written

\begin{equation}
    \gamma_{\text{e}} = n \sigma v \label{elasticrate}
\end{equation}

\noindent{where $n$ is the mean density, $\sigma$ is the
zero-energy s-wave cross-section, and $v$ is the mean relative
velocity. Assuming atoms are trapped in a state for which there is
no spin-exchange, there are three bad collisional processes:
background collisions, three-body recombination, and two-body
dipolar relaxation, with the per-atom loss rate respectively
denoted by $\gamma_{\text{bg}}$, $\gamma_{\text{3b}}$ and
$\gamma_{\text{2b}}$. Loss due to background collisions is to
first approximation independent of the cloud density and
temperature. (We examine the break-down of this approximation in
section \ref{heating} above.) We write the rate as}

\begin{subequations}\label{ratedefs}
    \begin{equation}\label{backgrd}
        \gamma_{\text{bg}} = \alpha
    \end{equation}

\noindent{where $\alpha$ is proportional to an effective total
residual pressure in the vacuum chamber.  The constant of
proportionality depends on the composition of the residual gas.
The other two rates are inelastic processes among the trapped
atoms. At low energies these rates become independent of collision
energy, and we can write}

    \begin{equation} \label{3body}
        \gamma_{\text{3b}} = \lambda n^2
    \end{equation}
    \begin{equation} \label{2body}
        \gamma_{\text{2b}} = \beta n
    \end{equation}
\end{subequations}

\noindent{where $\beta$ is the dipolar rate constant and $\lambda$
is the three-body rate constant.

The total rate of bad collisional processes is
$\gamma_{bad}=\gamma_{\text{bg}}+\gamma_{\text{3b}}+\gamma_{\text{2b}}$.
For good evaporation, the three inequalities,

\begin{subequations} \label{raterules}
    \begin{equation} \label{rule1}
        n\sigma v\gg\alpha
    \end{equation}
    \begin{equation} \label{rule2}
        n\sigma v\gg\lambda n^2
    \end{equation}
    \begin{equation}\label{rule3}
        n\sigma v\gg\beta n
    \end{equation}
\end{subequations}

\noindent must be separately satisfied. Moreover, the inequalities
must be satisfied not only initially, (\emph{i.e.} immediately
after the atoms are loaded into the trap) but also as evaporation
proceeds towards ever greater phase-space density, towards larger
$n$ and smaller $v$.

With respect to evaporating rubidium 87 and the lower hyperfine
level of sodium 23, nature has been kind. The respective values of
$\sigma$ are sufficiently high, and $\lambda$ and $\beta$
sufficiently low, that the collisional rate inequalities
\eqref{rule2} and \eqref{rule3} are routinely met along a
reasonable evaporation path through $n-T$ space to BEC. One need
``only" arrange for the initial trapped cloud to have sufficiently
large $n$, and design a vacuum chamber with sufficiently small
$\alpha$, and evaporation works. The main point of this section,
however, is that evaporation is likely to be possible even with
less favorable values of $\sigma$, $\beta$, and $\lambda$.  We
deal with the constraints \eqref{raterules} in order:

\subsubsection{Background loss} The technical difficulties of
ensuring that the elastic rate is much greater than the background
loss rate can be formidable. All the same, eq. \eqref{rule1} does
not represent a fundamental collisional limit, for the simple
reason that with sufficient laboratory effort $\alpha$ can be made
almost arbitrarily small. Lifetimes of $10^4$ s are attainable in
cryogenic magnetic traps. Cryogenic experiments are notoriously
difficult, but if one is willing to go to the effort and the
expense, one should ultimately succeed.  The challenge is only to
satisfy \eqref{rule1} at the beginning of evaporation. Once
efficient evaporation is established, $nv$ increases, while
$\alpha$ remains the same (except see section \ref{heating}
above.)

\subsubsection{Three-body recombination} \label{3bodyrules}
It is unlikely that $\lambda$ could ever be so large that
three-body recombination is a problem when the atoms are first
loaded into the trap.  As evaporation proceeds, however, density
increases and velocity decreases, such that inequality
\eqref{rule2} may well fail, if one is working with a species with
small $\sigma$ or large $\lambda$. The solution lies in
manipulating the trapping potential. Adiabatically ramping the
mean harmonic confinement frequency $\omega_{\text{t}}$ has no
effect on the phase-space density --- the ramp takes atoms neither
closer to nor further from the BEC transition. The mean density
and mean velocity are both affected by the ramp, however, such
that $\gamma_{\text{3b}}/\gamma_{\text{e}}=\lambda n^2/(n \sigma
v) \propto \omega_{\text{t}}$. Therefore, as long as one can
continue to turn down the confining strength of one's trap, one
can ensure that \eqref{rule2} remains satisfied all the way to the
BEC transition.

\subsubsection{Dipolar relaxation} \label{2bodyrules}
As with three-body recombination, if dipolar relaxation is to be a
problem, it will likely not be until late in the evaporative
process when the decrease in velocity may threaten the validity of
\eqref{rule3}. Adiabatically turning down the confining potential
is not helpful, because the scaling with $\omega_{\text{t}}$ has
the opposite sign: $\gamma_{\text{2b}}/\gamma_{\text{e}} \propto
{\omega_{\text{t}}}^{-1/2}$. Evidently one can win by turning
\emph{up} the trap. While this may be a useful tactic in a
particular laboratory situation, it is unsatisfactory in terms of
our ``Existence Proof" for evaporation. One may always make a trap
weaker, but there is likely to be an engineering-based upper limit
to trap strength, and if $\beta$ is particularly large, a problem
can remain. More fundamentally, if one is cursed with a species
for which both $\beta$ and $\lambda$ are large, there may be no
value of $\omega_{\text{t}}$ for which \eqref{rule2} and
\eqref{rule3} can be simultaneously satisfied in a cloud near
degeneracy.

Fortunately, one is not required to accept the value of $\beta$
nature provides. In fact, all one really has to do is to operate
the magnetic trap with a very low bias field. Below a threshold
field of perhaps five gauss, the value of $\beta$ for the lowest
hyperfine level drops rapidly to zero. This behavior is simple to
understand. At low temperature, the incoming collisional channel
must be purely s-wave.  Dipolar relaxation changes the projection
of spin angular momentum, so to conserve angular momentum the
outgoing collisional channel must be at d-wave or higher.  The
nonzero outgoing angular momentum means that there must be a
angular momentum barrier in the effective molecular potential, a
barrier which rises a few hundred microkelvin above the the
$r=\infty$ limit. This barrier is irrelevant for atoms trapped in
the stretch-state, for instance ($F=2, m_F=2$ in rubidium 87).
Dipolar relaxation releases the (relatively large) hyperfine
energy, and the outgoing atoms barely feel the the angular
momentum barrier. But if the atoms are trapped in the lower state
($F=1, m_F=-1$, in rubidium 87) the outgoing energy is only the
Zeeman energy in the trapping fields. For fields below about $5$
Gauss, this energy is insufficient to get the atoms back out over
the angular momentum barrier. If relaxation is to occur, it can
happen only at inter-atomic radii larger than outer turning-point
of the angular momentum barrier. For smaller and smaller fields,
the barrier gets pushed further out, with correspondingly lower
transition rates \cite{lowfieldsuppress}. This effect turns out to
be irrelevant in atomic hydrogen, since the lower hyperfine state
is $F=0$, and is not magnetically trappable. In most magnetically
trappable atoms and molecules, however, the lowest state has
$F\neq 0$, and one of the Zeeman sublevels will have no
spin-exchange, and suppressed dipolar relaxation.

To summarize this subsection, given (i) a modestly flexible
magnetic trap, (ii) an arbitrarily good vacuum, (iii) a true
ground-state with $F\neq 0$ and $S\neq 0$ and (iv) non-pathological
collisional properties, almost any species can be successfully
evaporated to BEC.

\subsection{Except of course \ldots } \label{exceptions}
The assumption of ``nonpathological properties" may actually be
fairly constraining.  For example: i) Rb-87 has a d-wave resonance
\cite{Boesten97}.  ii) Cs-133 appears to have a zero-field
Feshbach resonance \cite{Kokkelmans98}. iii) The elastic
cross-section of Rb-85 has a pronounced energy dependence and
drops to near-zero at a very inconvenient temperature
\cite{Courteille98, Roberts98}. One is forced to confront the fact
that at least in the alkalis, pathology seems almost to be the
rule. Also, for technical reasons having to do with low-frequency
electronics noise, it can be inconvenient to trap atoms at fields
below about $0.5$ G, which may not always be low enough to
completely suppress dipolar relaxation. Also the scaling arguments
discussed above assume T-independent elastic cross-section;
optical molasses temperatures are not always low enough to ensure
that one is in this regime.  Further, we have completely neglected
the issue of heating. Therefore we invite you to adopt an
appropriately skeptical attitude towards the universality of the
arguments outlined in this section.

\subsection{But on the other hand \ldots }
Requiring that a particular atomic or molecular species be able,
on its own, to cool to BEC is unnecessarily restrictive.
Sympathetic cooling has been shown to work well.  As far as we
know, thorough modeling of the collisional requirements for
sympathetic cooling has not been performed, but simple estimates
suggest that in the presence of a robustly evaporating working
fluid, the collisional requirements on the species to be cooled,
and on interspecies collisions, are down by a factor of ten or
more from those outlined in the first paragraph of section
\ref{evapcriteria}.

Once you have cooled to the BEC transition, your ability to
accumulate a significant number of atoms in the condensate depends
on the scattering length's being positive. Here again, even in
total ignorance, your odds are pretty good: see the argument in
the footnote \cite{potentialthoughts}.

\acknowledgments We acknowledge support from the National Science
Foundation, the Office of Naval Research, and the National
Institute of Standards and Technology.  We have benefited
enormously from ongoing discussions with other members of the JILA
BEC collaboration.


\begin{thebibliography}{100}
\bibitem{Hehistorynote}
Actually, the notion that superfluidity was a manifestation of
Bose-Einstein
  Condensation was seen as highly controversial in the early theory of liquid
  Helium. Landau rejected the idea in favor of quantum hydrodynamics. See the
  ref. {\cite{Gavroglu95}} for more discussion.
\bibitem{Cornell98c}
While it has been argued that such systems are not true mixtures,
the
  distinction turns out to be largely academic, except in special cases.
\bibitem{vortexphase}
Only slightly more complicated! If we assume the vortex has
winding number
  {$n$} and that its core lies along the axis of symmetry of a condensate, and
  if we use cylindrical coordinates {$x_i = (r_i,\theta_i,z_i)$}, then {$\delta
  \phi(x_1,x_2) = n(\theta_1 - \theta_2)$}.
\bibitem{Oortexists}
The existence of hot, diffuse cloud of trapped atoms has been
discussed in
  refs. {\cite{Mewes96a, Stamper-Kurn98a, Burt97, Myattthesis}}.
\bibitem{lifetimenote}
Usually, but not always. Unless care is taken in the experimental
set-up,
  lifetimes can be limited by absorption of stray scattered laser light, or by
  unwanted radio-frequency magnetic noise.
\end{thebibliography}

\end{document}